\documentclass{article}

\usepackage{arxiv}

\usepackage[utf8]{inputenc} 
\usepackage[T1]{fontenc}    
\usepackage{hyperref}       
\usepackage{url}            
\usepackage{booktabs}       
\usepackage{amsfonts}       
\usepackage{nicefrac}       
\usepackage{microtype}      
\usepackage{float}
\usepackage{graphicx}
\usepackage{caption}
\usepackage{subcaption}
\usepackage{natbib}
\usepackage{doi}
\usepackage{siunitx}
\usepackage{mathtools}
\usepackage{natbib}
\usepackage{wrapfig}
\usepackage{multicol}
\usepackage[dvipsnames]{xcolor}
\usepackage{tikz}
\definecolor{ao(english)}{rgb}{0.0, 0.5, 0.0}

\title{Cycle-skipping mitigation using misfit measurements based on differentiable dynamic time warping}

\author{ {Fuqiang~Chen, Daniel Peter, Matteo Ravasi}\\
\\
	Physical Sciences and Engineering (PSE)\\
	King Abdullah University of Science and Technology\\
	Thuwal, Jeddah 23599 \\
	\\
	\texttt{fuqiang.chen@kaust.edu.sa} \\
}

\renewcommand{\shorttitle}{FWI with differentiable DTW misfit}

\graphicspath{{./Figs/}}
\begin{document}
\maketitle

\begin{abstract}
The dynamic time warping (DTW) misfit function has been used for wave-equation inversion to mitigate the local minima issue. However, the original DTW distance is not smooth; therefore it can yield a strong discontinuity in the adjoint source. Such a weakness does not help nonlinear inverse problems converge to a plausible minimum by any means. We introduce in geophysics the smooth DTW misfit function, which has demonstrated its performance in time series classification, clustering, and prediction. The fundamental idea of the smooth DTW misfit measurement is to replace the min operator with its smooth relaxation. This replacement makes the warping distance differentiable. Moreover, considering that the optimal warping plan is an indicator of the traveltime difference between the observed and synthetic trace, we can construct a penalization term based on it such that the misfit measured by the penalized differentiable DTW distance is weighted in favor of the traveltime difference. Numerical examples demonstrate the advantage of the penalized differentiable DTW misfit function over the conventional
non-differentiable one.
\end{abstract}


\section{Introduction}
In large-scale nonlinear inverse problems, gradient-based iterative solvers suffer from the local minima issue. As a result, if a given initial estimate is far from the global minimum, the final estimate may likely be far from it too. Unfortunately until nowadays this pitfall still cannot be avoided in the application of full-waveform inversion (FWI)---a typical example of nonlinear inverse problems. This obstacle keeps FWI from making dramatic changes in the seismic imaging industry \cite[]{symesmvafwi08,symes2020}. An ideal scenario reaching a plausible local minimum is to start with low-frequency data, for example, below $\SI{3.0}{\hertz}$. However, those frequencies are generally not recorded in practice due to instrument response and the contamination from strong noise and surface-related multiples; this is especially true for reflection data. Alternatively, a successful application of FWI requires the availability of a necessarily good initial model from a previous step of tomography \cite[]{tomo10} or migration velocity analysis \cite[]{savamva04,symesmvafwi08}. As a rule of thumb, if the phase discrepancy between the observed and synthetic data from a given initial model is less than one half-cycle, then FWI using the least-squares misfit function can converge to a plausible local minimum \cite[]{VirieuxFWIoverview}. Otherwise, the inversion will converge to an implausible one.
This is a very strict requirement though. 

Besides improving the availability of data and the initial model, many attempts have also been made to construct misfit functions in which informative local minima are approachable for gradient-based optimization methods. A long-established and productive research topic for this purpose is to construct novel misfit functions of traveltime difference. The rationale behind this strategy is that misfit functions focusing on traveltime difference measurements show the chance to rid the implausible local minima and they can break the limitation of cycle-skipping. If recalling the industrial-strength methods \cite[]{tomo10} used for seismic velocity model estimation, human-controlled workflows guarantee that the misfit function calculations in grid or model reflection tomography are entirely traveltime oriented. The automatic estimation of traveltime differences between two datasets for FWI is significantly difficult. Various misfit functions based on cross-correlation or deconvolution have been developed for this purpose \cite[]{LuoSchuster1991,LuoSava2011,WarnerGuasch2016,ZhuFomel2016}. Traveltime difference estimated by cross-correlation or deconvolution lacks robustness and still requires proper data selection. \cite{self2021} proposed combining the local cross-correlation with dynamic time warping to moderate this restriction.
 
Other misfit functions, e.g., those based on the optimal transport distance \cite[]{engquist2016optimal,Metivier2016,Yang2018,self.1DMongeRadon,self.Sinkhorn,Jaimeetal2019} have also been proposed to mitigate the local minima issue for FWI. The transport plan from Monge or Kantorovich formulation \cite[]{self.Sinkhorn} can be used as an indicator of traveltime difference. This property makes optimal transport distance possible to mitigate the local minima issue. \cite{engquist2016optimal} and \cite{Metivier2016} empirically showed that the optimal transport distance presents an expanded convex zone compared to the misfit function based on $\mathcal{L}^2$-norm. However, the derivative of optimal transport distance based on Kantorovich formulation with synthetic data is not numerically accurate or it is impractical to obtain the accurate derivative in a finite number of iterations \cite[]{self.Sinkhorn}. The derivative of optimal transport distance based on 1D Monge formulation \cite[]{engquist2016optimal,Yang2018,self.1DMongeRadon} can be calculated analytically; however, data transformation introduced to adapt seismic data to probability density functions will degrade the convex property of optimal transport distance based on Monge formulation \cite[]{engquist2016optimal,Yang2018,self.1DMongeRadon}. 
   
The dynamic time warping (DTW) distance \cite[]{Sakoe_DTW1978} is another measurement that can indicate the traveltime difference. It is widely used for applications that seek to evaluate the similarity between images or sequences even with different shapes \cite[]{Cuturi2017SoftDTWAD,MenschBlondel2018,Guen2019ShapeAT}. In seismic data processing, DTW is used to estimate time-shifts between datasets \cite[]{Hale2013DTW} and it has demonstrated its robustness when compared to cross-correlation methods. In the geophysical community, \cite{Yang2014UsingIW} and \cite{MaHale2013} have proposed the misfit function based on the DTW distance to mitigate the local minima issue in waveform inversion. As we demonstrate later, the original DTW distance \cite[]{Sakoe_DTW1978,Hale2013DTW,Yang2014UsingIW} is not smooth and results in sharp changes in the adjoint sources \cite[]{selfsdtw2021}. We argue that these issues will affect the efficiency of seismic model evaluation by an iterative approach. 

The smooth or differentiable DTW distance replaces the min operator in the original DTW distance with its differentiable alternative \cite[]{Cuturi2017SoftDTWAD}. \cite{Guen2019ShapeAT} proposed a penalized differentiable DTW distance by the warping path to further enhance the temporal discrepancy in misfit measurements. The temporal discrepancy has another name in the seismic inversion regime by traveltime difference. We introduce such a penalized differentiable DTW misfit function in this paper and apply it to wave-equation inversion. We first use a two-parameter inverse problem to show that the penalized differentiable DTW misfit function can mitigate the local minima issue. Then we use the Chevron synthetic data to discuss how waveform inversion based on the introduced misfit can outperform inversion based on the original DTW distance.

\section{Theory}
Given seismic traces $f=\{f_1,\cdots,f_{nt}\}$ and $g=\{g_1,\cdots,g_{nt}\}$, the DTW distance \cite[]{Sakoe_DTW1978} is the minimum accumulation cost along a $np$-node path parametrized by $(i_k,j_k)$. It can be defined as:
\begin{equation}
\varepsilon_0(f,g):=\min_{(i_k,j_k)}\displaystyle\sum_{k=1}^{np} \mathcal{D}_{i_k,j_k},
\label{eq:dtw0}
\end{equation} 
where $\mathcal{D}_{i,j}$ represents the discrepancy between $f_i$ and $g_j$ and it can be defined by different methods, such as the local cross-correlation \cite[]{self2021} and the squared Euclidean cost $\mathcal{D}_{i,j}=(f_i-g_j)^2$ \cite[]{Sakoe_DTW1978, Hale2013DTW, Cuturi2017SoftDTWAD,MenschBlondel2018,Guen2019ShapeAT}, the one we take in this paper. 
A greedy algorithm to find this warping path for practical problems is intractable. \cite{Sakoe_DTW1978} proposed the dynamic programming algorithm to solve equation \ref{eq:dtw0}. The algorithm divides the overall problem into subproblems in which the solution to the subproblem can be found by a sorting process with the $\min$ operator. The output of this $\min$ operator is however globally nondifferentiable. Locally, it changes as a step function, which implies that the DTW distance in equation \ref{eq:dtw0} is not even continuous. We still use the word ``derivative'' to stand for the change of DTW distance caused by a perturbation in synthetic data. The approximated derivative information of $\varepsilon_0(f,g)$ has already been used to retrieve a model to reduce the data misfit or image defocussing \cite[]{MaHale2013,Yang2014UsingIW}. This inaccuracy at least will slow down the iteration process if it does not ruin the final result. 

\cite{Cuturi2017SoftDTWAD} proposed an equivalent definition of DTW distance:
\begin{equation}
\varepsilon_\gamma({f},{g}):={\min}^{\gamma}\big\{\big<\mathcal{W}, \mathcal{D} \big>,{\mathcal{W}}\in \mathbb{R}^{nt\times nt}\big\},
\label{eq:soft-dtw}
\end{equation} 
where the matrix $\mathcal{W}$ represents a feasible warping plan, $\big<\mathcal{W},\mathcal{D}\big>=\sum_{i,j}\mathcal{W}_{i,j}\mathcal{D}_{i,j}$ denotes the Frobenius inner product, and 
\begin{equation}
{\min}^{\gamma}(x_1,\cdots,x_n):=-\gamma\log\displaystyle\sum_{i=1}^n e^{-x_i/\gamma} \label{eq:softmin_a}.
\end{equation}
Equation \ref{eq:softmin_a} is also known as the differentiable or smooth minimization operator. Because the min operator is replaced by its smooth version, the resulted DTW distance in equation \ref{eq:soft-dtw} becomes differentiable. Equation \ref{eq:softmin_a} can cause the overflow issue. An overflow-free formulation is
\begin{equation}
{\min}^{\gamma}(x_1,\cdots,x_n):=x'-\gamma \log\displaystyle\sum_{i=1}^{n}e^{-\frac{x_i-x'}{\gamma}},\label{eq:sofasdtmin_b}
\end{equation}
where $x'=\min(x_1,\cdots,x_n)$. 
Equation $\ref{eq:soft-dtw}$ recovers the non-differentiable DTW distance admitted by equation \ref{eq:dtw0} if $\gamma=0$ \cite[]{Cuturi2017SoftDTWAD}. We restrict the warping path $\mathcal{W}$ by, for example, only allowing it to arise following three directions: $\rightarrow$, $\downarrow$ and $\searrow$ \cite[]{Cuturi2017SoftDTWAD}. This restriction eliminates the arbitrary warping schemes. \cite{Sakoe_DTW1978} developed a fast algorithm to calculate the DTW distance in equation \ref{eq:dtw0}, which is also suitable for the calculation of differentiable DTW distance in equation \ref{eq:soft-dtw}. It consists of a sweeping process with the boundary conditions $\mathcal{A}[:,0]=+\infty$, $\mathcal{A}[0,:]=+\infty$, and $\mathcal{A}[0,0]=0$:
\begin{equation}
\mathcal{A}_{i,j}=\mathcal{D}_{i,j}+{\min}^{\gamma} \big( 
            \mathcal{A}_{i-1,j-1},
            \mathcal{A}_{i-1,j},
            \mathcal{A}_{i,j-1}\big),
            \label{eq:f-dtw-accum}
\end{equation}
where $\mathcal{A}$ stands for the accumulation array memorizing all paths satisfying the forward marching directions. Plunging the definition of $\min^{\gamma}$, we have 
\begin{equation}
\mathcal{A}_{i,j}=\mathcal{D}_{i,j}-\gamma\log\displaystyle\sum\limits_{(i',j')\in \mathbb{P}} e^{-\mathcal{A}_{i+i',j+j'}/\gamma},
\label{eq:f-dtw-accum-logsumexp}
\end{equation}
where $\mathbb{P}=\{(-1,0),(0,-1),(-1,-1)\}$. To get the DTW distance, we evaluate equation \ref{eq:f-dtw-accum-logsumexp} from $\mathcal{A}_{1,1}$ to $\mathcal{A}_{nt,nt}$. We can call this step forward sweeping. After the forward sweeping, we have $\varepsilon_\gamma({f},{g})=\mathcal{A}_{nt,nt}$ \cite[]{Cuturi2017SoftDTWAD,MenschBlondel2018}. The differentiable DTW distance $\varepsilon_\gamma({f},{g})$ can reflect the traveltime difference. This capacity can be further enhanced by introducing the optimal warping plan as a penalty term \cite[]{Guen2019ShapeAT}. Then DTW distance with penalization is defined as:
\begin{subequations}
\begin{align}
\varepsilon_{\gamma,\lambda}({f},{g};\mathcal{I}):=&\varepsilon_\gamma + \lambda \big<\mathcal{W}^{*},\mathcal{I}\big> \\
=& \varepsilon_\gamma +\lambda\phi,
\end{align}
\end{subequations}
where $\mathcal{W}^{*}$ denotes the optimal warping plan, and $\mathcal{I}$ represents the prior distribution of certain warping path, for example, $\mathcal{I}_{i,j}=(i-j)^2/nt^2$ \cite[]{Guen2019ShapeAT} or $\mathcal{I}=\mathcal{D}$ \cite[]{pmlr-v130-blondel21a}. We will call $\gamma$ the smoothness parameter and $\lambda$ the penalization parameter.
In the following, we explain the algorithms developed by \cite{Cuturi2017SoftDTWAD} and \cite{MenschBlondel2018}, respectively to efficiently calculate the derivative of $\varepsilon_{\gamma}$ and $\phi$ w.r.t. $f$. Considering $\varepsilon_\gamma=\mathcal{A}_{nt,nt}$ and the dependency of $\mathcal{A}_{nt,nt}$ on all $\mathcal{A}_{i,j}$, a straightforward derivative calculation will soon become impractical with the increase in dimension of $\mathcal{A}$. \cite{Cuturi2017SoftDTWAD} discovered a recursive formula to calculate $\frac{\partial \varepsilon_\gamma}{\partial \mathcal{A}_{i,j}}$. Given the 3-point forward sweeping template in equation \ref{eq:f-dtw-accum-logsumexp}, we can have
\begin{equation}
\frac{\partial \varepsilon_\gamma}{\partial \mathcal{A}_{i,j}}=\displaystyle\sum_{(i',j')\in\mathbb{C}}{\frac{\partial \varepsilon_\gamma}{\partial \mathcal{A}_{i+i',j+j'}}\frac{ \partial \mathcal{A}_{i+i',j+j'}}{\partial \mathcal{A}_{i,j}}},
\label{eq:deda}
\end{equation}
where $\mathbb{C}=\{(1,0),(0,1),(1,1)\}$. Let $\mathcal{X}_{i,j}=\frac{\partial \varepsilon_\gamma}{\partial \mathcal{A}_{i,j}}$ and plug it into equation \ref{eq:deda}, then we have
\begin{equation}
\mathcal{X}_{i,j}=\displaystyle\sum_{(i',j')\in\mathbb{C}}\mathcal{X}_{i+i',j+j'}\frac{\partial \mathcal{A}_{i+i',j+j'}}{\partial \mathcal{A}_{i,j}}.
\label{eq:EE}
\end{equation}
Based on equation \ref{eq:f-dtw-accum-logsumexp}, we can have 
\begin{equation}
\frac{\partial \mathcal{A}_{i+1,j}}{\partial \mathcal{A}_{i,j}}=\frac{e^{-\mathcal{A}_{i,j}/\gamma}}{\displaystyle\sum\limits_{(i',j')\in \mathbb{P}}e^{-\mathcal{A}_{i+1+i',j+j'}/\gamma}}.
\label{eq:D2D_1}
\end{equation}
Taking the logarithm of both sides of equation \ref{eq:D2D_1} and according to equation \ref{eq:f-dtw-accum-logsumexp}, we obtain
\begin{equation}
\gamma\log \frac{\partial \mathcal{A}_{i+1,j}}{\partial \mathcal{A}_{i,j}}=\mathcal{A}_{i+1,j}-\mathcal{A}_{i,j}-\mathcal{D}_{i+1,j}.
\label{eq:gamma1}
\end{equation}
Similarly, we can have
\begin{subequations}
\begin{align}
\gamma\log \frac{\partial \mathcal{A}_{i,j+1}}{\partial \mathcal{A}_{i,j}}&=\mathcal{A}_{i,j+1}-\mathcal{A}_{i,j}-\mathcal{D}_{i,j+1},\\
\gamma\log \frac{\partial \mathcal{A}_{i+1,j+1}}{\partial \mathcal{A}_{i,i}}&=\mathcal{A}_{i+1,i+1}-\mathcal{A}_{i,j}-\mathcal{D}_{i+1,j+1}.
\end{align}
\label{eq:gamma2}
\end{subequations}
Combining equations \ref{eq:EE}, \ref{eq:gamma1}, \ref{eq:gamma2} and $\mathcal{X}_{nt,nt}=\frac{\partial \mathcal{A}_{nt,nt}}{\partial \mathcal{A}_{nt,nt}}=1$, we can fill the matrix $\mathcal{X}$, which takes a backward sweeping process. In summary, we obtain the differentiable DTW distance by one forward sweeping and the optimal warping path $\mathcal{W}^{*}=\mathcal{X}$ by a backward sweeping \cite[]{Cuturi2017SoftDTWAD}. Then the derivative of differentiable DTW distance $\varepsilon_\gamma$ w.r.t. synthetic data $f$ can be calculated as: 
\begin{equation}
\frac{\partial \varepsilon_{\gamma}}{\partial f_i}=2\displaystyle\sum\limits_{j=1}^{nt}\frac{\partial \varepsilon_{\gamma}}{\partial \mathcal{A}_{i,j}}(f_i-g_j).
\label{eq:adj_0}
\end{equation}

Calculating the derivative of the penalization term $\phi$ w.r.t synthetic data involves the Hessian of DTW distance with $\mathcal{A}_{i,j}$ on all feasible paths. The direct calculation is prohibitive. Fortunately, we only need the Hessian-vector product, which is equivalent to the derivative of the directional derivative of $\varepsilon_\gamma$ as shown below:
\begin{subequations}
\begin{align}
\frac{\partial \phi}{\partial \mathcal{A}_{k,l}}&=\displaystyle\sum\limits_{i=1}^{nt}\sum\limits_{j=1}^{nt}\frac{\partial \mathcal{X}_{i,j}}{\partial \mathcal{A}_{k,l}}\mathcal{I}_{i,j},\label{eq:depsilonD4tensor}\\
&=\underbrace{\sum\limits_{i=1}^{nt}\sum\limits_{j=1}^{nt}\frac{\partial^2 \varepsilon_\gamma}{\partial \mathcal{A}_{k,l}\partial \mathcal{A}_{i,j}}\mathcal{I}_{i,j}}_{\textrm{Hessian product}},\label{eq:hessian}\\
&=\displaystyle\frac{\partial}{\partial \mathcal{A}_{k,l}}\underbrace{\sum\limits_{i=1}^{nt}\sum\limits_{j=1}^{nt}\frac{\partial \varepsilon_\gamma}{\partial \mathcal{A}_{i,j}}\mathcal{I}_{i,j}}_{\textrm{directional derivative}}.\label{eq:hessiantodirectionalder}
\end{align}
\end{subequations}
Compared to equation \ref{eq:hessian}, equation \ref{eq:hessiantodirectionalder} converts the Hessian product into the first derivative of the directional derivative of differentiable DTW distance along $\mathcal{I}$. The directional derivative is an immediate result considering $\partial \epsilon_\gamma/\partial \mathcal{A}_{i,j}$ is already available. But it is still not clear how to then efficiently calculate its derivative with $\mathcal{A}_{k,l}$ in equation \ref{eq:hessiantodirectionalder}. \cite{MenschBlondel2018} developed a recursive algorithm to calculate the directional derivative as well as the derivative of directional derivative. To explain this algorithm, we first introduce
\begin{equation}
\dot{\mathcal{X}}_{m,n}=\sum\limits_{i=1}^{m}\sum\limits_{j=1}^{n}\frac{\partial \mathcal{A}_{m,n}}{\partial \mathcal{A}_{i,j}}\mathcal{I}_{i,j} ,
\end{equation}
as the directional derivative of warping cost $\mathcal{A}_{m,n}$ for subproblems. Based on the sweeping template in equation \ref{eq:f-dtw-accum}, $\mathcal{A}_{m,n}$ is directly dependent on $\mathcal{A}_{m+i,n+j}$ with $(i,j)\in \mathbb{P}$, this leads to a recursive calculation for $\dot{\mathcal{X}}_{m,n}$ as
\begin{equation}
\displaystyle\dot{\mathcal{X}}_{m,n}=\mathcal{I}_{m,n}+\displaystyle\sum\limits_{(m',n')\in \mathbb{P}}\dot{\mathcal{X}}_{m+m',n+n'}\frac{\partial \mathcal{A}_{m,n}}{\partial \mathcal{A}_{m+m',n+n'}}.
\label{eq:directionalderivativeforall}
\end{equation} 
Equation \ref{eq:directionalderivativeforall} represents a forward recursion with the boundary conditions $\dot{\mathcal{X}}[0,:]=0$ and $\dot{\mathcal{X}}[:,0]=0$. Finally, the directional derivative of differentiable DTW distance $\varepsilon_\gamma$ along $\mathcal{I}$ is $\dot{\mathcal{X}}_{nt,nt}=<\mathcal{X},\mathcal{I}>$. 

Calculating ${\partial \dot{\mathcal{X}}_{nt,nt}}/{\partial \mathcal{A}_{i,j}}$ exactly follows the same procedure as we calculate ${\partial \varepsilon_\gamma}/{\partial \mathcal{A}_{i,j}}$. A small change in $\mathcal{A}_{i,j}$ directly causes a change in $\mathcal{A}_{i+i',j+j'}$ and $\dot{\mathcal{X}}_{i+i',j+j'}$ for $(i',j')\in\mathbb{C}$. Then wen have 
\begin{equation}
{\color{red}\frac{\partial \dot{\mathcal{X}}_{nt,nt}}{\partial \mathcal{A}_{i,j}}}=\displaystyle\sum\limits_{(i',j')\in\mathbb{C}}{\color{violet}\frac{\partial \dot{\mathcal{X}}_{nt,nt}}{\partial {\mathcal{A}_{i+i',j+j'}}}}{\color{blue}\frac{\partial {\mathcal{A}_{i+i',j+j'}}}{\partial {\mathcal{A}_{i,j}}}}+\sum\limits_{(i',j')\in\mathbb{C}}{\color{brown}\frac{\partial\dot{\mathcal{X}}_{nt,nt}}{\partial {\dot{\mathcal{X}}_{i+i',j+j'}}}}{\color{ao(english)}\frac{\partial{\dot{\mathcal{X}}_{i+i',j+j'}}}{\partial {\mathcal{A}_{i,j}}}},
\label{eq:dofdd}
\end{equation} 
where five terms are colored differently for a convenient explanation \cite[]{MenschBlondel2018}. The {\color{violet}first term} on the right-hand side of equation \ref{eq:dofdd} implies a backward recursive formulation to calculate {\color{red}${\partial \dot{\mathcal{X}}_{nt,nt}}/{\partial {\mathcal{A}_{i,j}}}$}. The {\color{blue}second term} has already been calculated when solving ${\partial \varepsilon_\gamma}/{\partial {\mathcal{A}_{i,j}}}$. The result of the {\color{brown}third term} can be derived as follows: Based on equation \ref{eq:directionalderivativeforall}, 
\begin{subequations}
\begin{align}
\frac{\partial \dot{\mathcal{X}}_{m,n}}{\partial {\dot{\mathcal{X}}_{i,j}}}&=\sum\limits_{(i',j')\in\mathbb{C}}\frac{\partial\dot{\mathcal{X}}_{m,n}}{\partial {\dot{\mathcal{X}}_{i+i',j+j'}}}\frac{\partial{\dot{\mathcal{X}}_{i+i',j+j'}}}{\partial {\dot{\mathcal{X}}_{i,j}}}, \\
&=\sum\limits_{(i',j')\in\mathbb{C}}\frac{\partial\dot{\mathcal{X}}_{m,n}}{\partial {\dot{\mathcal{X}}_{i+i',j+j'}}}\frac{\partial{\dot{\mathcal{A}}_{i+i',j+j'}}}{\partial {\dot{\mathcal{A}}_{i,j}}}.
\label{eq:thirdt}
\end{align}
\end{subequations}
Comparing equation \ref{eq:thirdt} and \ref{eq:EE}, we have 
\begin{equation}
\frac{\partial \dot{\mathcal{X}}_{nt,nt}}{\partial {\dot{\mathcal{X}}_{i,j}}}=\mathcal{X}_{i,j}.
\label{eq:aliasforW}
\end{equation}
Finally, the {\color{ao(english)}last term} on the right of equation \ref{eq:dofdd} can be calculated by differentiating equation \ref{eq:directionalderivativeforall} for $(i',j')\in\mathbb{C}$, then we obtain
\begin{equation}
\frac{\partial \dot{\mathcal{X}}_{i+i',j+j'}}{\partial {\mathcal{A}_{i,j}}}=\displaystyle \sum\limits_{(m',n')\in \mathbb{P}}\dot{\mathcal{X}}_{i+i'+m',j+j'+n'} \frac{\partial^2 \mathcal{A}_{i+i',j+j'}}{\partial \mathcal{A}_{i+i'+m',j+j'+n'}\partial \mathcal{A}_{i,j}},
\label{eq:hessianofmin}
\end{equation}
The derivative term in Equation \ref{eq:hessianofmin} represents the Hessian of the differentiable $\min$ operator. With $\partial \varepsilon_\gamma/\partial \mathcal{A}_{i,j}$ and $\partial \dot{\mathcal{X}}_{nt,nt}/\partial \mathcal{A}_{i,j}$ ready, the adjoint source from penalized differentiable DTW distance can be calculated as 
\begin{equation}
\frac{\partial \varepsilon_{\gamma,\lambda}}{\partial f_i}=2\displaystyle\sum\limits_{j=1}^{nt}\bigg(\frac{\partial \varepsilon_{\gamma}}{\partial \mathcal{A}_{i,j}}+\lambda\frac{\partial \dot{\mathcal{X}}_{nt,nt}}{\partial \mathcal{A}_{i,j}}\bigg)(f_i-g_j),
\label{eq:adj}
\end{equation}
if $\mathcal{D}$ is defined by the squared Euclidean cost and $\mathcal{I}$ is independent on $f$.

\section{Examples}
\begin{figure}[H]
     \centering
    \begin{subfigure}[htbp]{0.372\textwidth}
        \begin{tikzpicture}
        \node[inner sep=0pt] (duck) at (0,0)
        {\includegraphics[width=\textwidth]{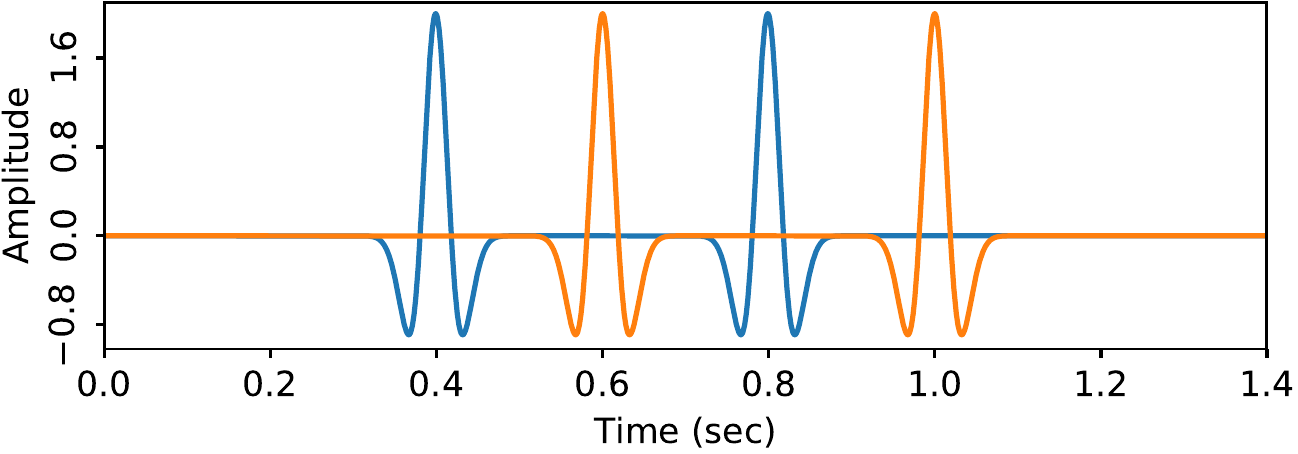}};
        \node[align=center,fill=white,draw] at (-1.7,.35) {(a)};
        \end{tikzpicture}
    \end{subfigure}
    \begin{subfigure}[htbp]{0.372\textwidth}
        \begin{tikzpicture}
        \node[inner sep=0pt] (duck) at (0,0)
        {\includegraphics[width=\textwidth]{obs_0.600_1.000_syn_0.399_0.799}};
        \node[align=center,fill=white,draw] at (-1.7,0.35) {(b)};
        \end{tikzpicture}
    \end{subfigure}
      \caption{\label{fig:rickers} Observed trace (orange) and two instances of synthetic traces (blue).}
\end{figure}

\begin{figure}[H]
    \centering
    \includegraphics[width=.3\textwidth]{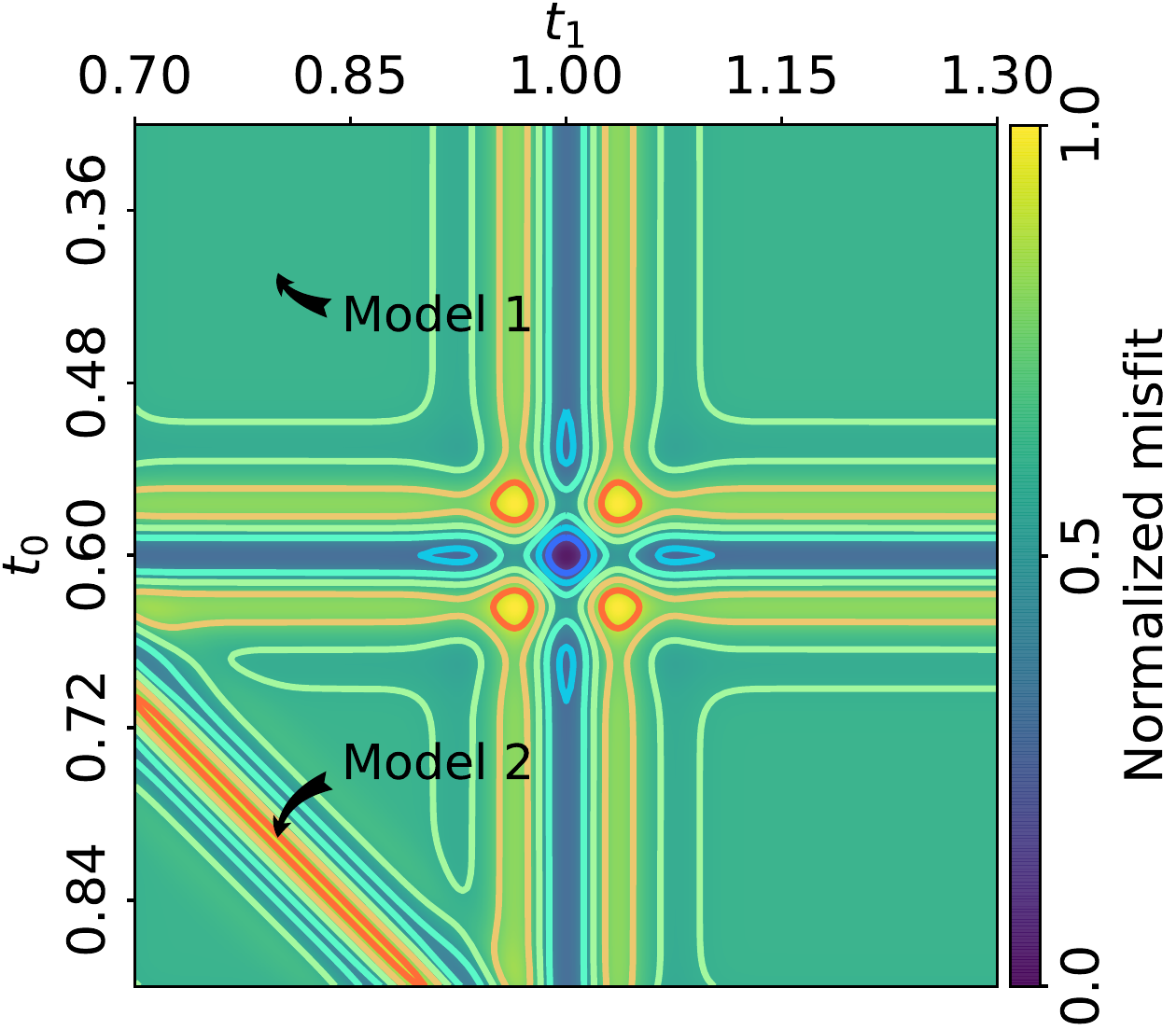}
    \caption{\label{fig:l2_msf} The map of normalized misfit measurements based on $\mathcal{L}^2$-norm. The model 1 and model 2 correspond to the models for the synthetic data in Figure \ref{fig:rickers}a and \ref{fig:rickers}b, respectively.}
\end{figure}

\begin{figure}[H]
     \centering
     \begin{subfigure}[b]{0.33\textwidth}
         \centering
         \includegraphics[width=\textwidth]{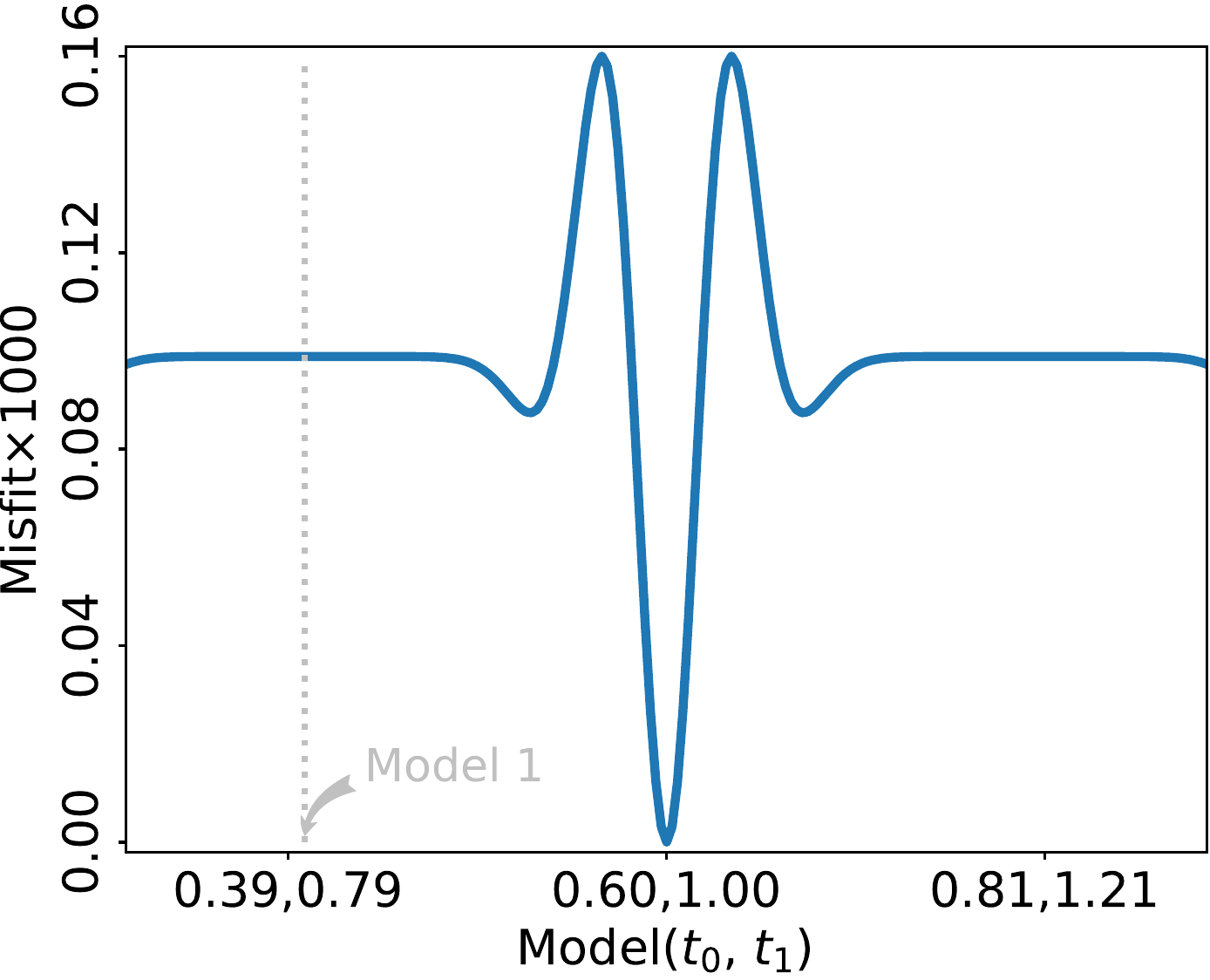}
         \caption{}
         \label{fig:y equals x}
     \end{subfigure}
     \begin{subfigure}[b]{0.33\textwidth}
         \centering
         \includegraphics[width=\textwidth]{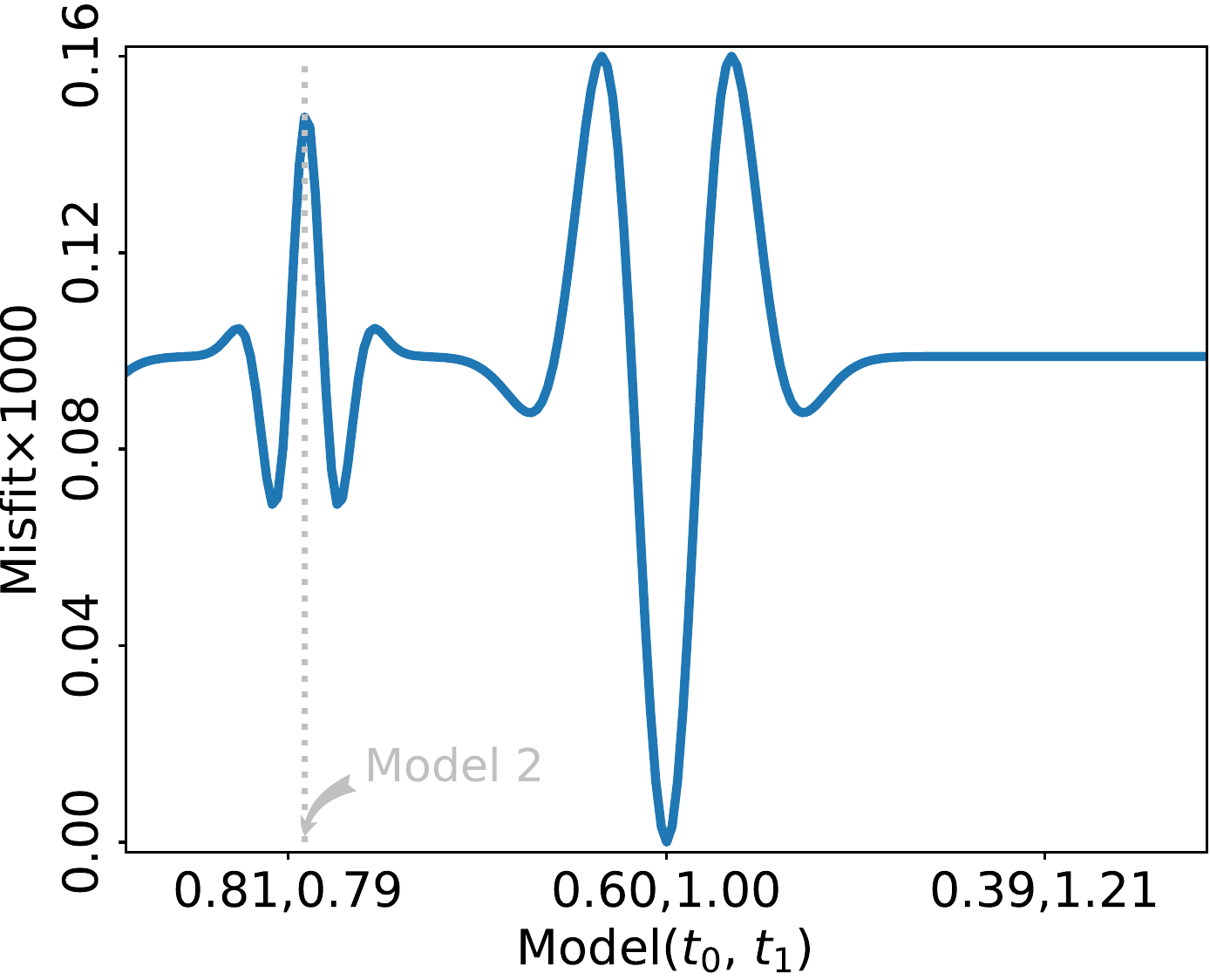}
         \caption{}
         \label{fig:three sin x}
     \end{subfigure}
      \caption{\label{fig:details_l2}(a) the diagonal and (b) anti-diagonal details of misfit map in Figure \ref{fig:l2_msf}.}
\end{figure}

We begin by testing the property of the introduced misfit measurements with a two-parameter inverse problem \citep{self2021}. The governing equation is as follows:
\begin{equation*}
f(t)=r(t; t_0, f_m)+r(t; t_1, f_m),
\end{equation*} 
where $r(t; t_0, f_m)$ denotes the Ricker wavelet with the peak frequency $f_m$ and the peak amplitude at time $t_0$. Figure \ref{fig:rickers} shows the observed and two instances of synthetic data, where the peak frequency $f_m=\SI{12.0}{\hertz}$. The misfit measurement based on $\mathcal{L}^2$-norm is illustrated in Figure \ref{fig:l2_msf}, where the black arrows indicate the models that generate the synthetic traces in Figure \ref{fig:rickers}. Figure \ref{fig:details_l2} shows the details along the diagonal and antidiagonal of the misfit map in Figure \ref{fig:l2_msf}. The results in Figure \ref{fig:l2_msf} and Figure \ref{fig:details_l2} will be a reference to the results from the penalized differentiable DTW misfit. 
\begin{multicols}{2}
\end{multicols}

\begin{figure}[H]
     \centering
     \begin{subfigure}[b]{0.23\textwidth}
         \centering
         \includegraphics[width=\textwidth]{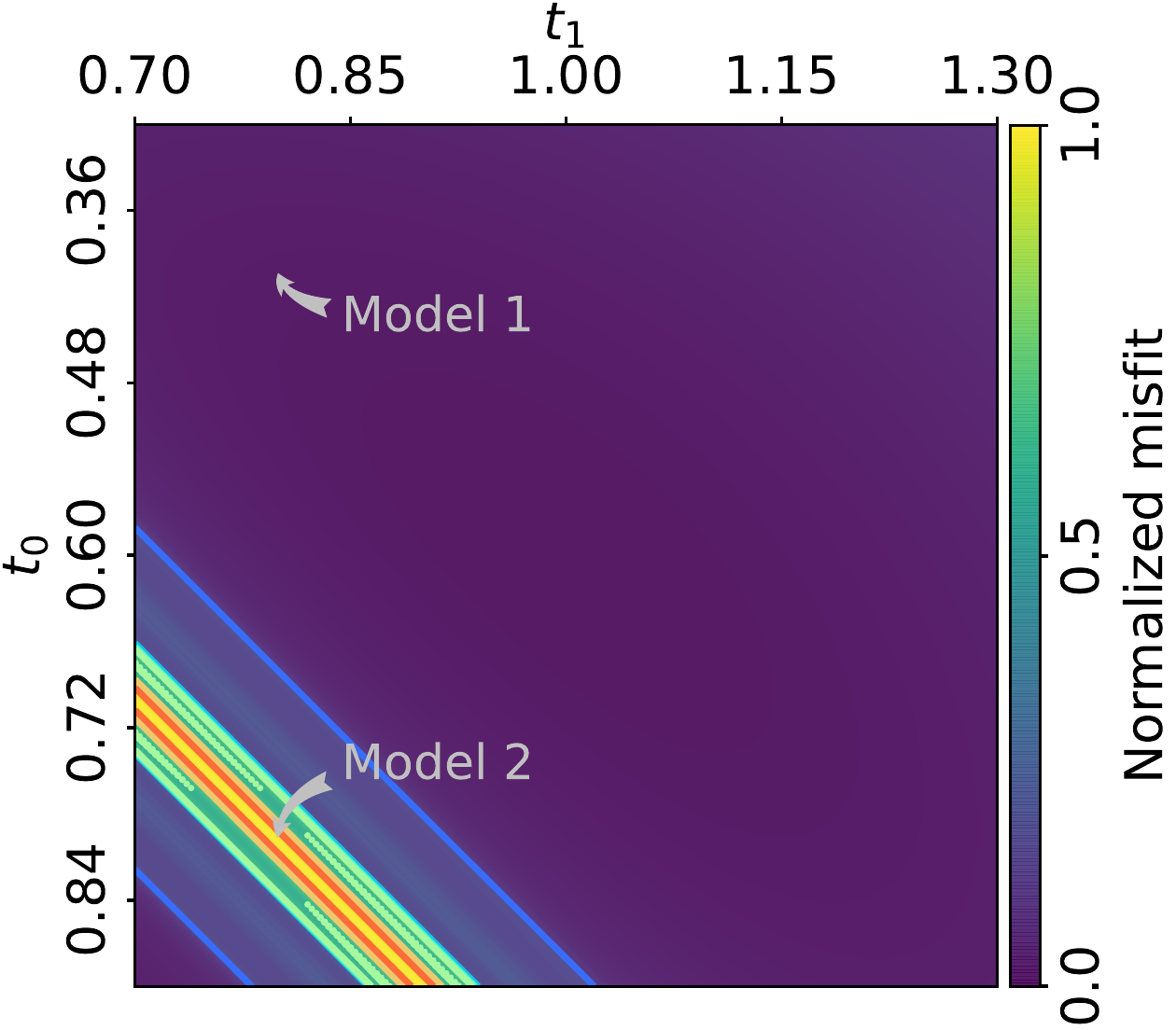}
         \caption{}
         \label{fig:y equals x}
     \end{subfigure}
     \begin{subfigure}[b]{0.23\textwidth}
         \centering
         \includegraphics[width=\textwidth]{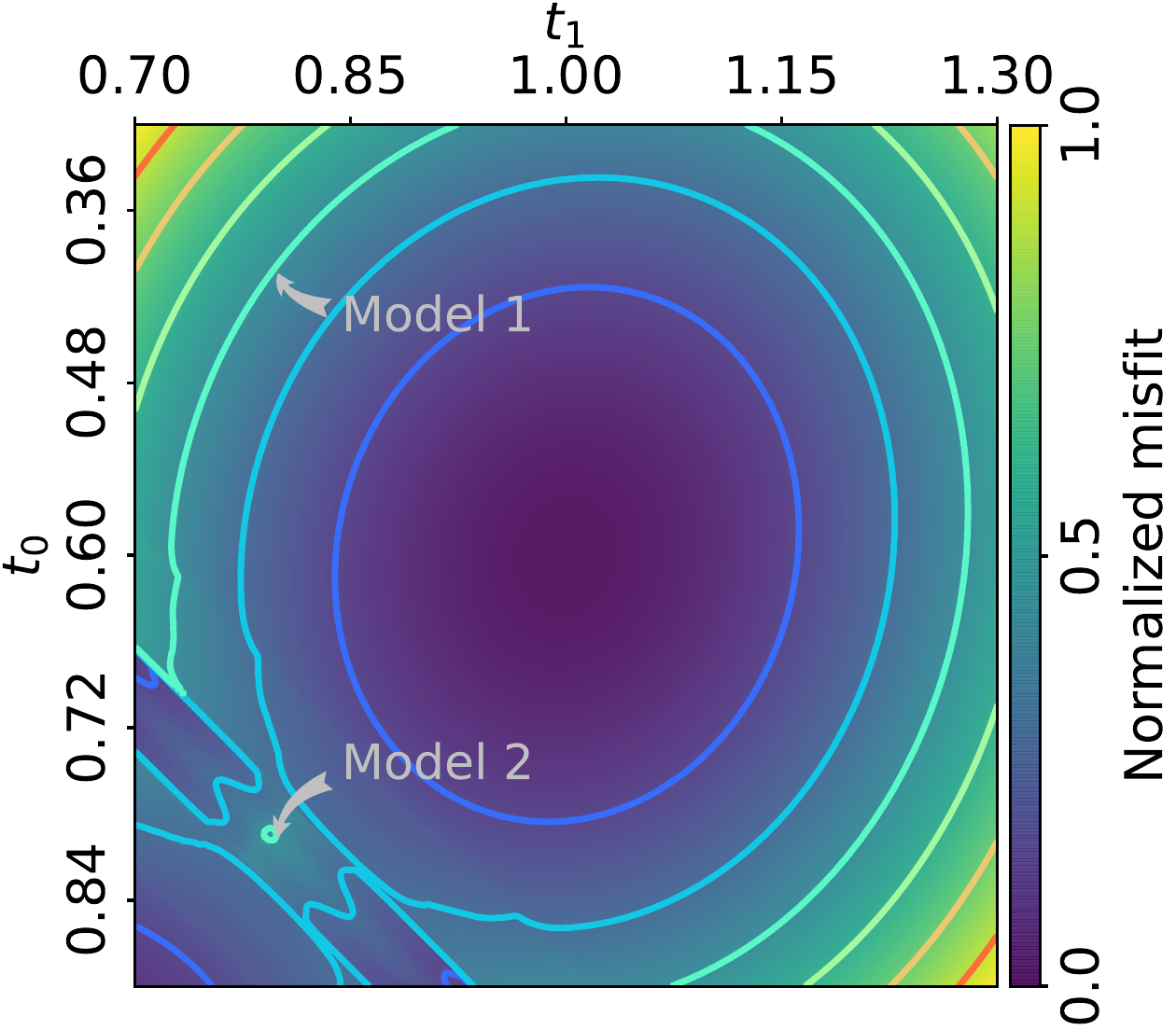}
         \caption{}
         \label{fig:three sin x}
     \end{subfigure}
          \begin{subfigure}[b]{0.23\textwidth}
         \centering
         \includegraphics[width=\textwidth]{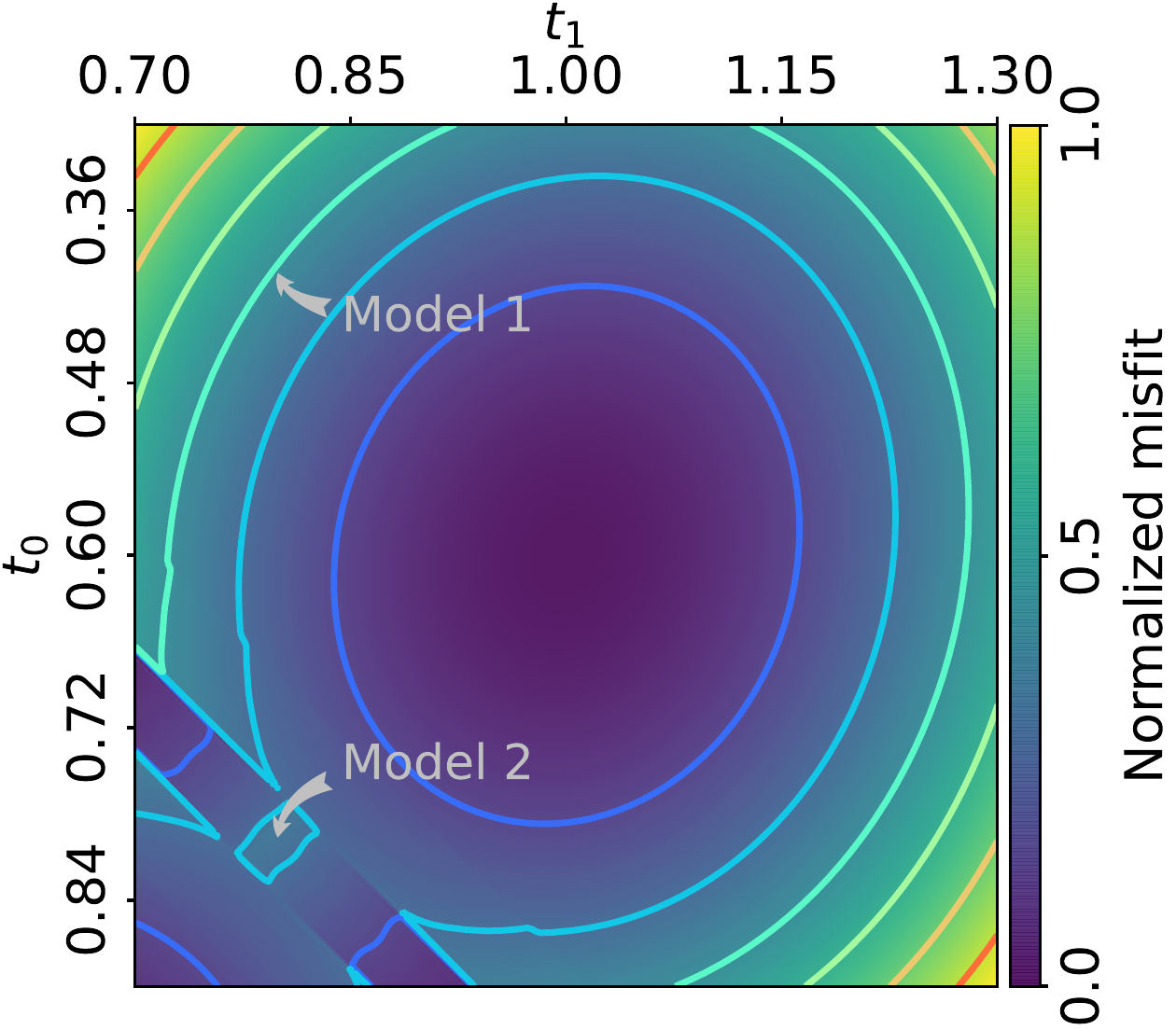}
         \caption{}
         \label{fig:y equals x}
     \end{subfigure}
     
     \begin{subfigure}[b]{0.23\textwidth}
         \centering
         \includegraphics[width=\textwidth]{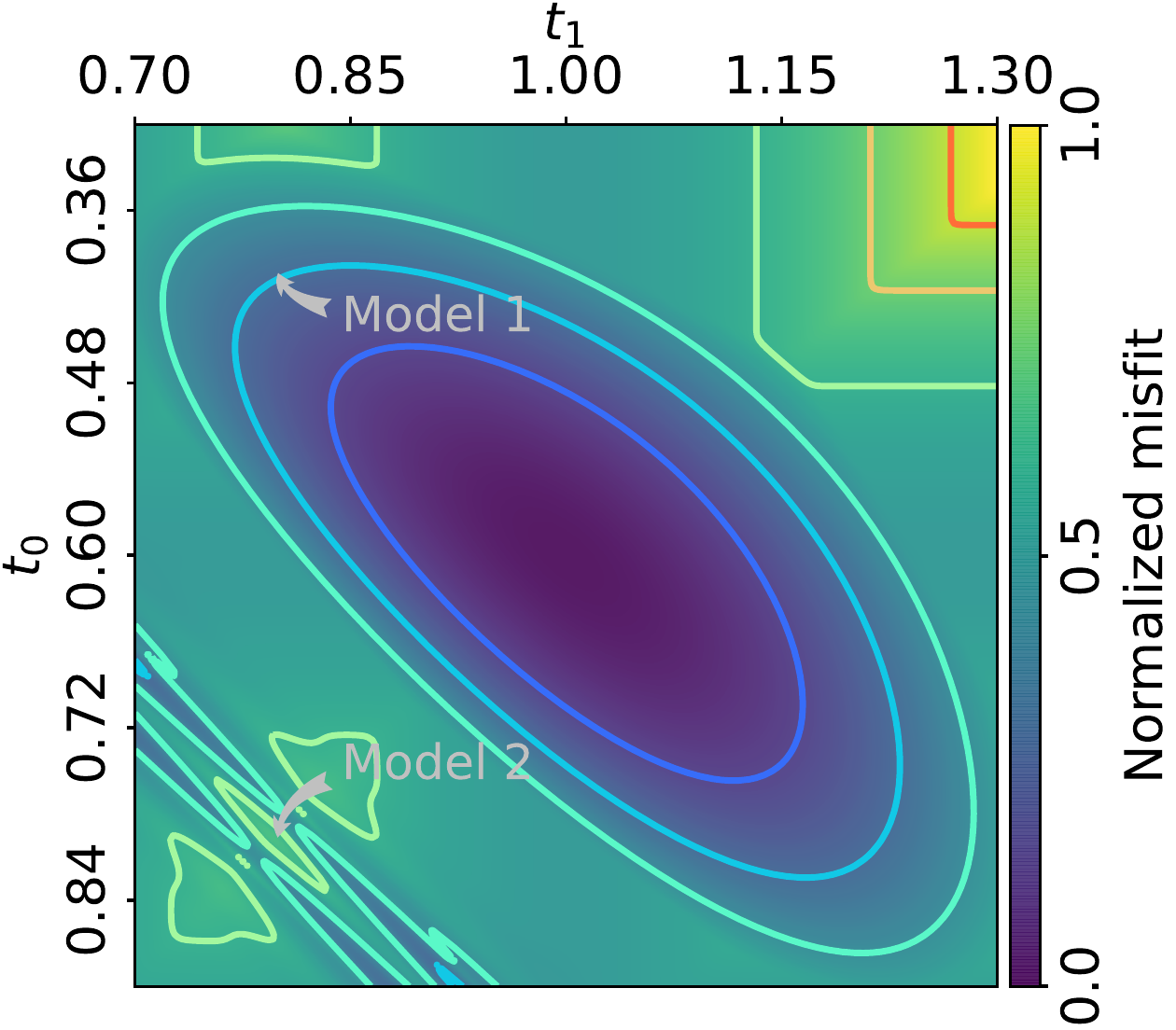}
         \caption{}
         \label{fig:three sin x}
     \end{subfigure}
          \begin{subfigure}[b]{0.23\textwidth}
         \centering
         \includegraphics[width=\textwidth]{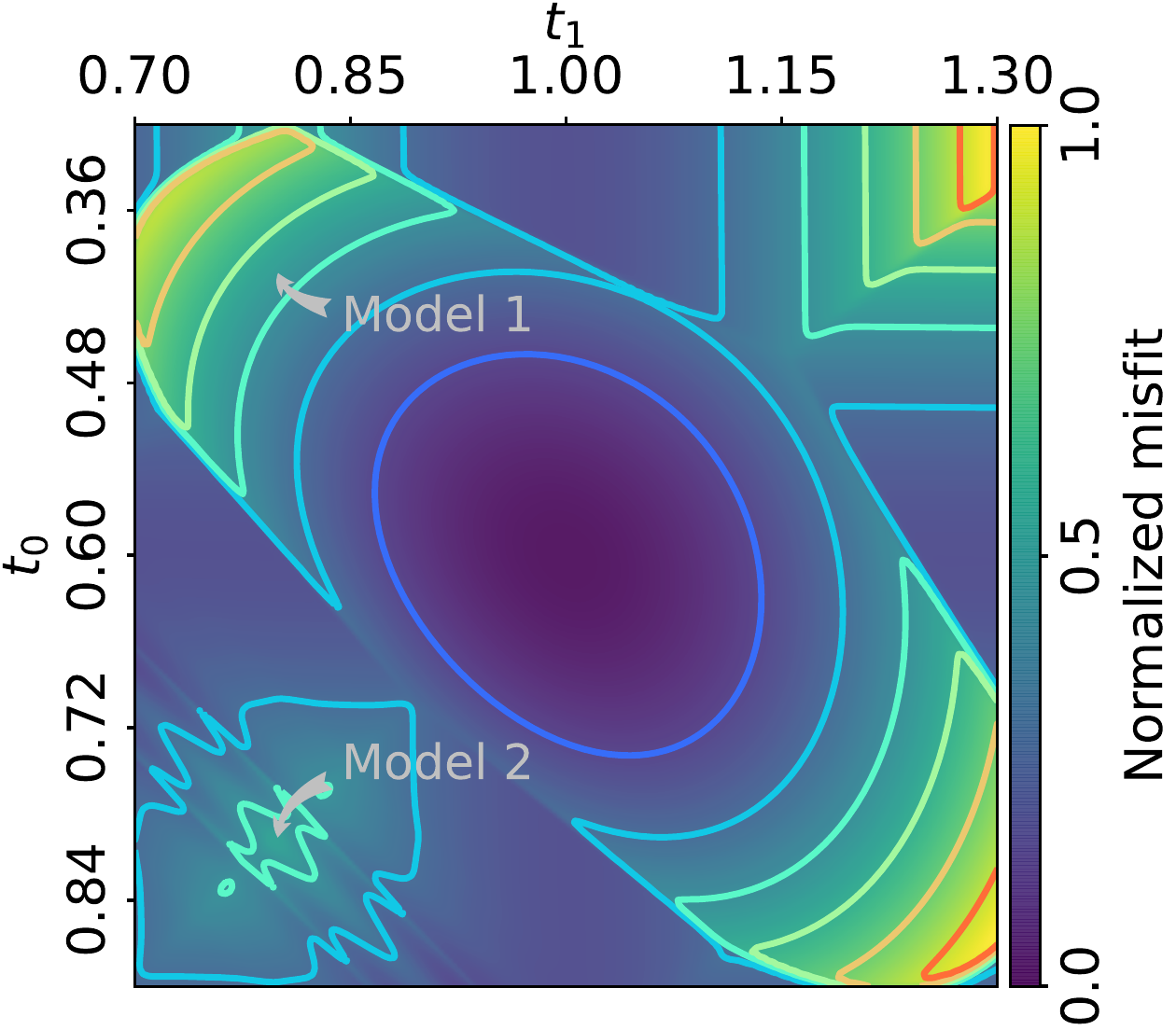}
         \caption{}
         \label{fig:y equals x}
     \end{subfigure}
     \begin{subfigure}[b]{0.23\textwidth}
         \centering
         \includegraphics[width=\textwidth]{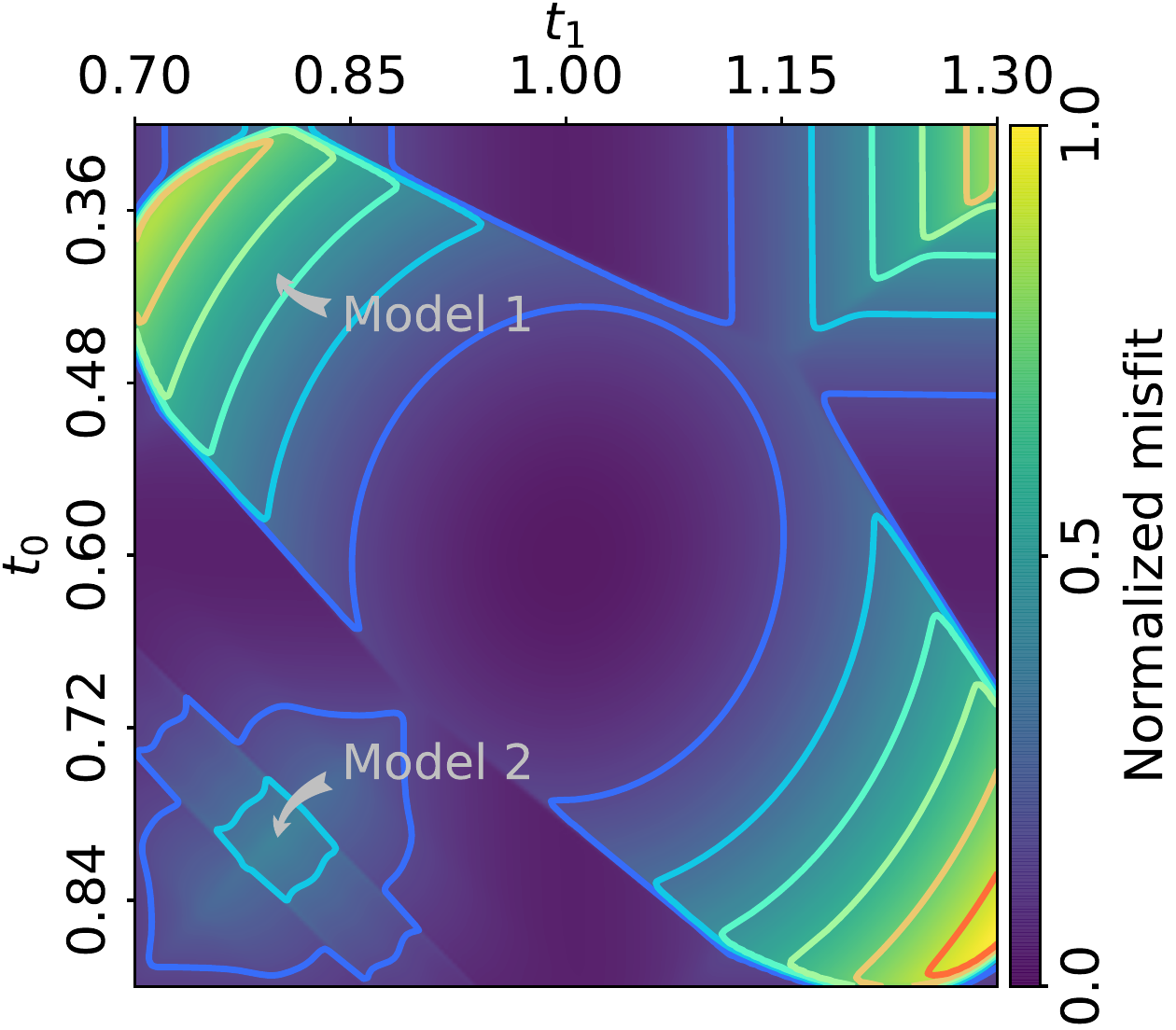}
         \caption{}
         \label{fig:three sin x}
     \end{subfigure}
     
     \begin{subfigure}[b]{0.23\textwidth}
         \centering
         \includegraphics[width=\textwidth]{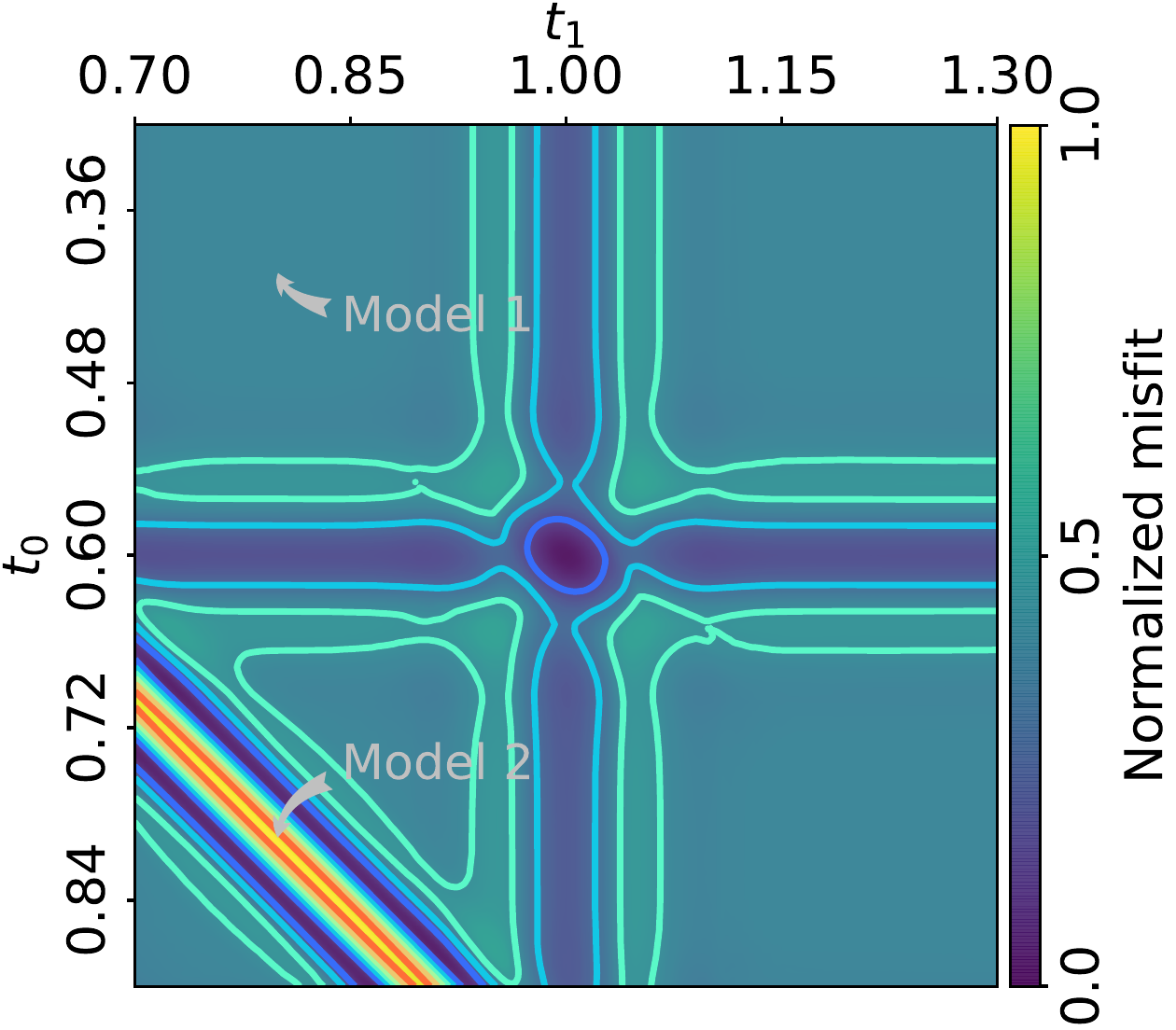}
         \caption{}
         \label{fig:y equals x}
     \end{subfigure}
     \begin{subfigure}[b]{0.23\textwidth}
         \centering
         \includegraphics[width=\textwidth]{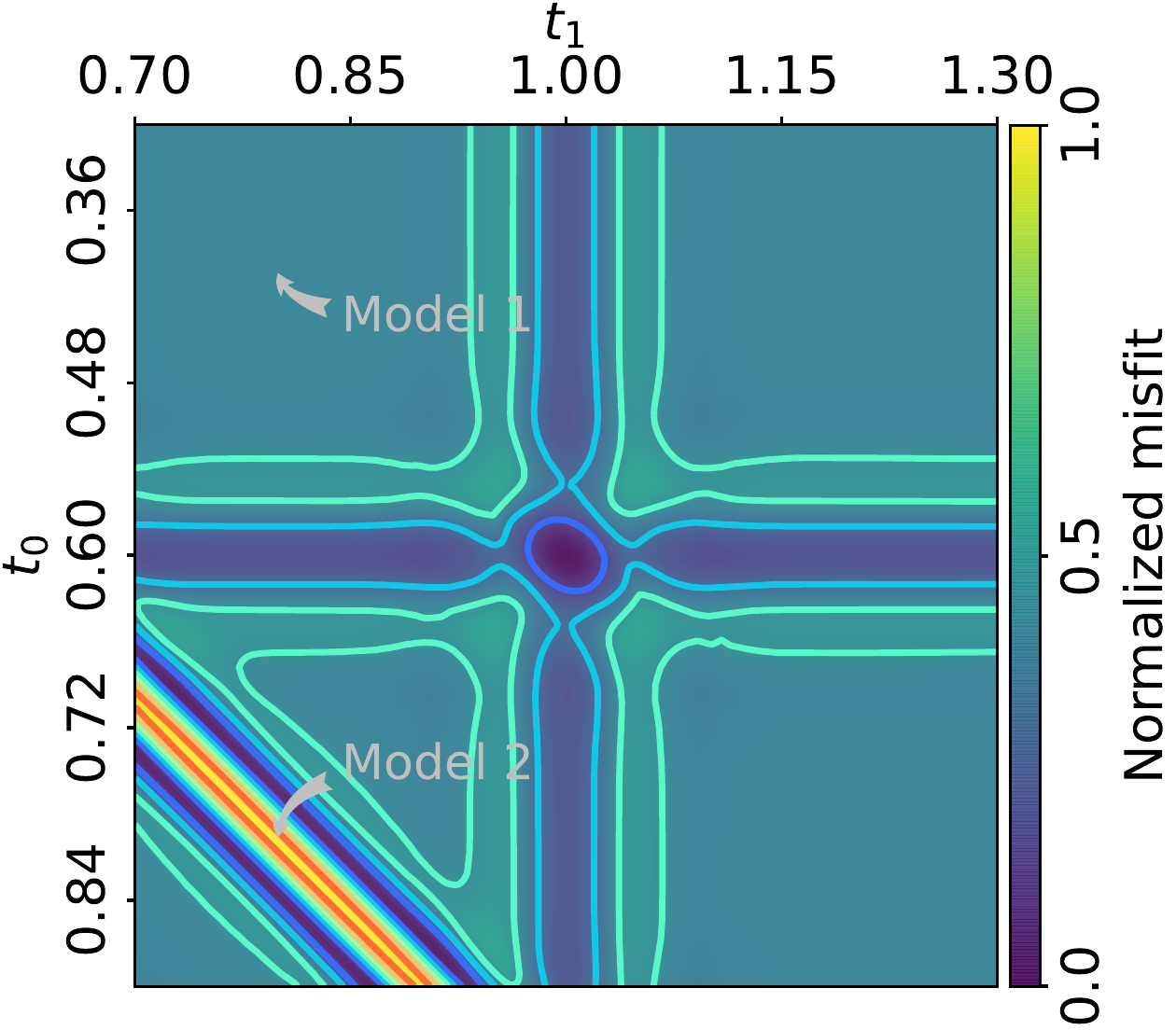}
         \caption{}
         \label{fig:three sin x}
     \end{subfigure}
     \begin{subfigure}[b]{0.23\textwidth}
         \centering
         \includegraphics[width=\textwidth]{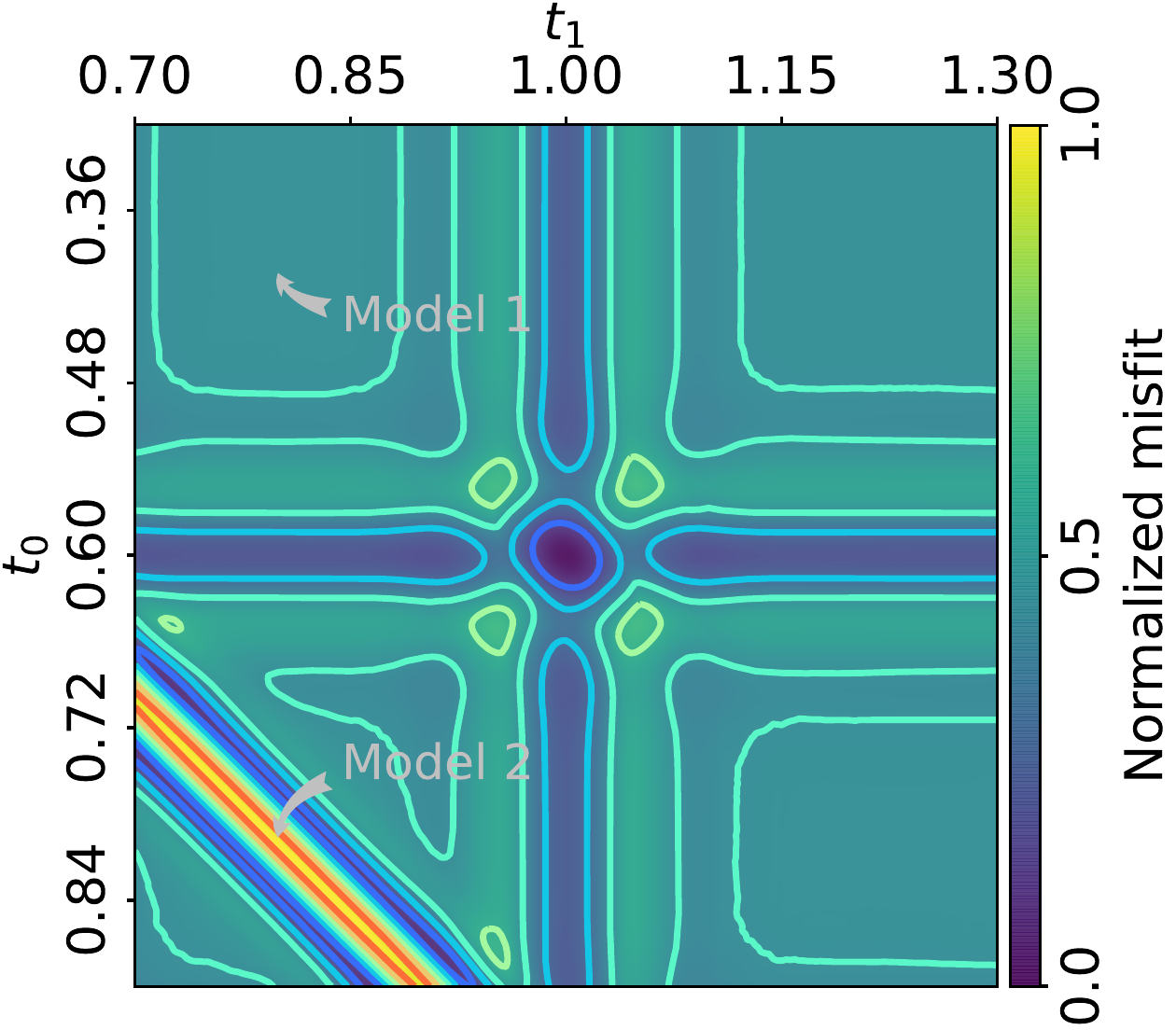}
         \caption{}
         \label{fig:three sin x}
     \end{subfigure}
      \caption{\label{fig:dtw_msf} Misfit measured by penalized differentiable DTW distance $\varepsilon_{\gamma,\lambda}$ with different values of hyper-parameters $\gamma$ and $\lambda$, where $\gamma={[0.01,1,100]}$ from top to bottom and $\lambda={[0, 9, 99]}$ from left to right. Every misfit map is normalized in ${[0,1]}$ by doing $\varepsilon$ $\leftarrow$ $\varepsilon$-$\varepsilon$.min() followed by $\varepsilon$ $\leftarrow$ $\varepsilon$/$\varepsilon$.max(). The model 1 and model 2 correspond to the models for the synthetic data in Figure \ref{fig:rickers}a and \ref{fig:rickers}b, respectively. }
\end{figure}
\begin{figure}[H]
     \centering
     \begin{subfigure}[b]{0.23\textwidth}
         \centering
         \includegraphics[width=\textwidth]{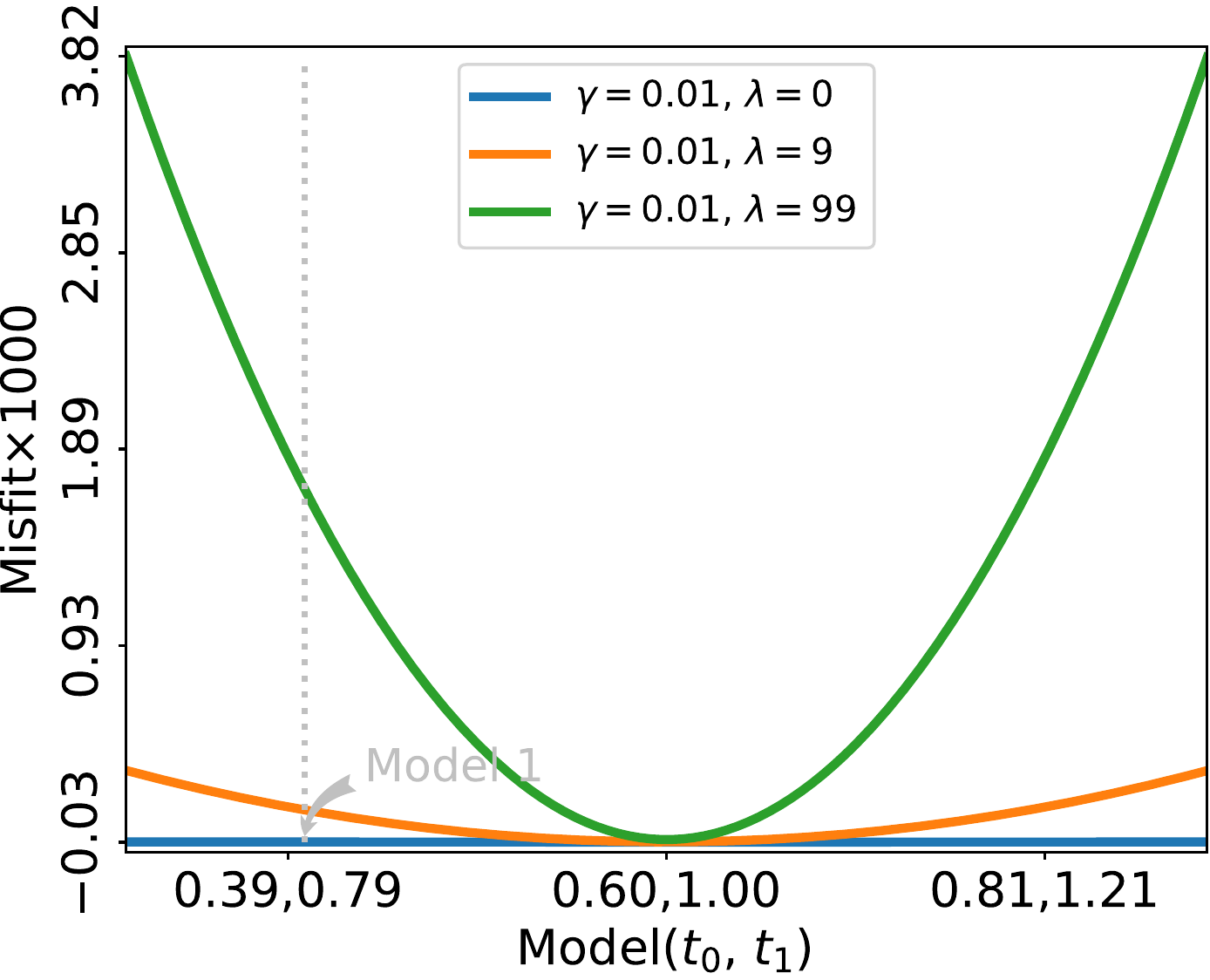}
         \caption{}
         \label{fig:y equals x}
     \end{subfigure}
     \begin{subfigure}[b]{0.23\textwidth}
         \centering
         \includegraphics[width=\textwidth]{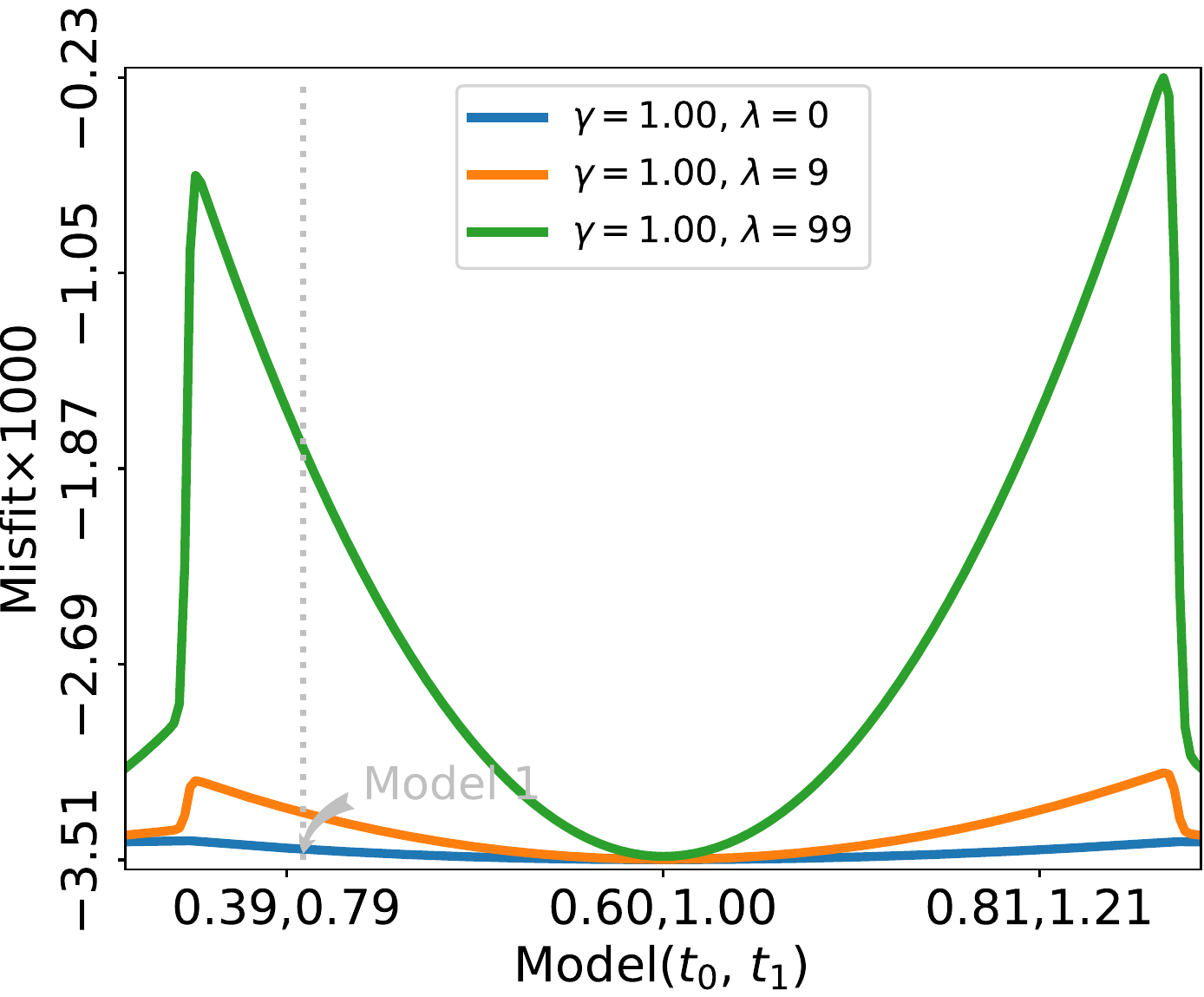}
         \caption{}
         \label{fig:three sin x}
     \end{subfigure}
          \begin{subfigure}[b]{0.23\textwidth}
         \centering
         \includegraphics[width=\textwidth]{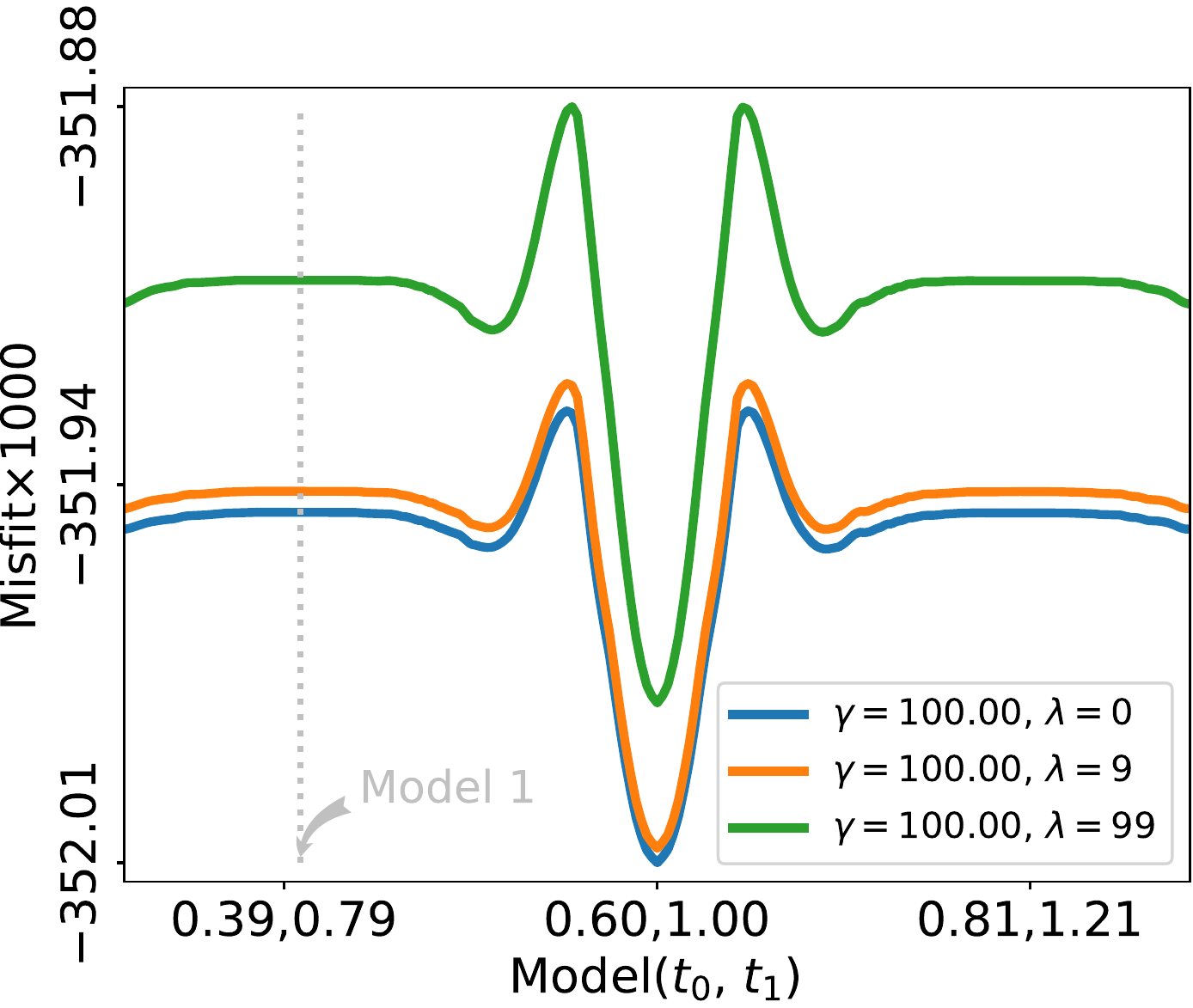}
         \caption{}
         \label{fig:y equals x}
     \end{subfigure}
     
     \begin{subfigure}[b]{0.23\textwidth}
         \centering
         \includegraphics[width=\textwidth]{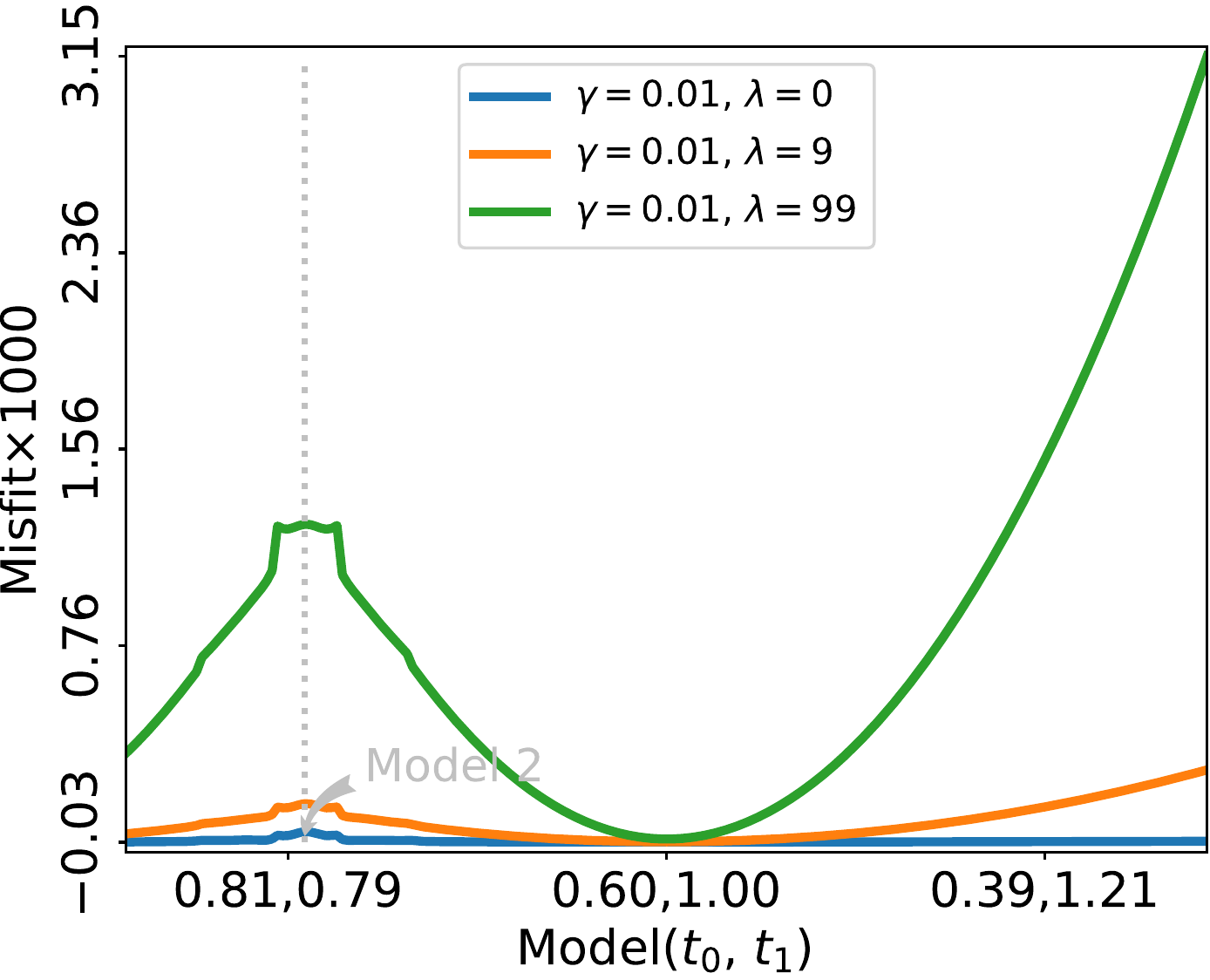}
         \caption{}
         \label{fig:three sin x}
     \end{subfigure}
          \begin{subfigure}[b]{0.23\textwidth}
         \centering
         \includegraphics[width=\textwidth]{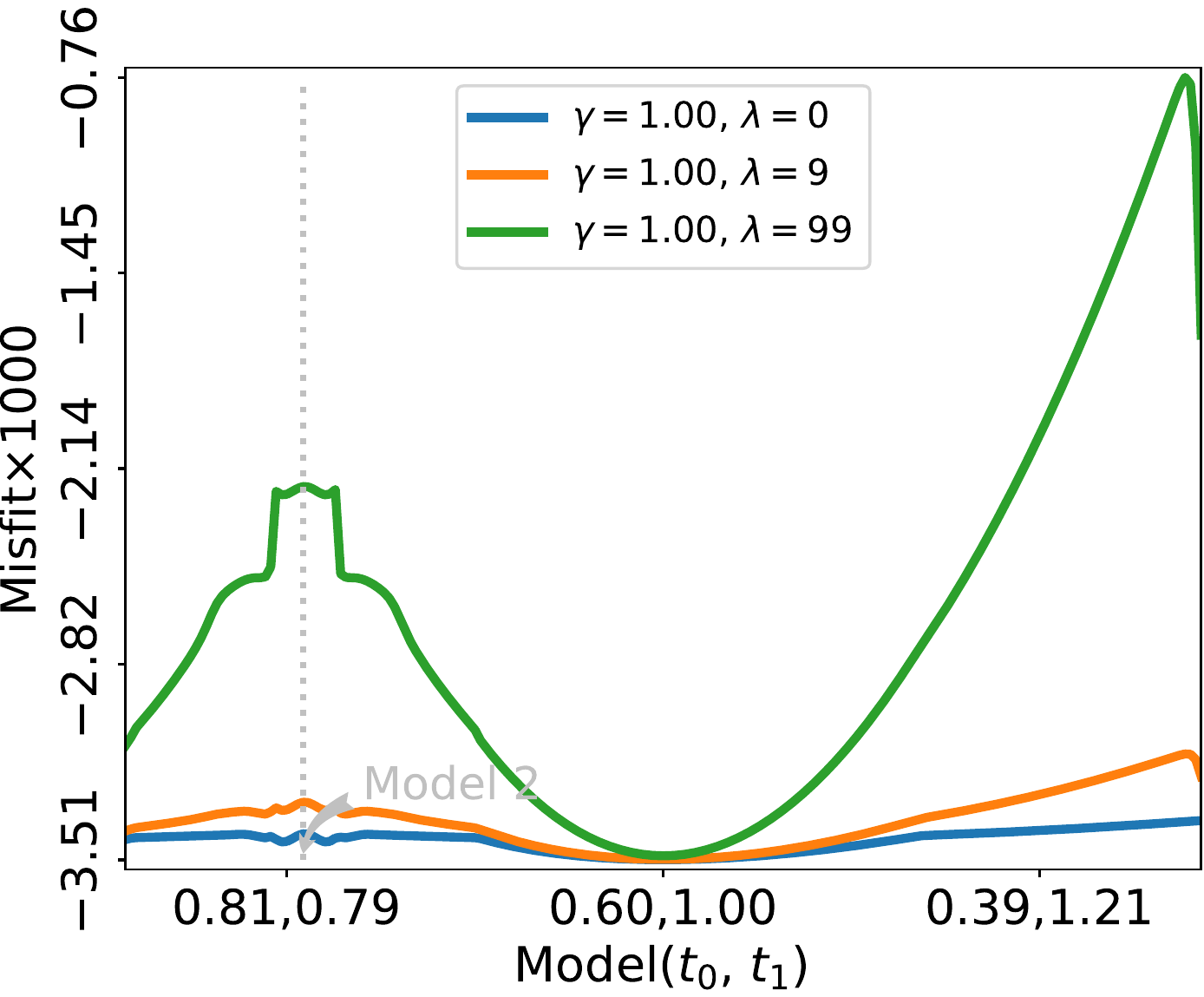}
         \caption{}
         \label{fig:y equals x}
     \end{subfigure}
     \begin{subfigure}[b]{0.23\textwidth}
         \centering
         \includegraphics[width=\textwidth]{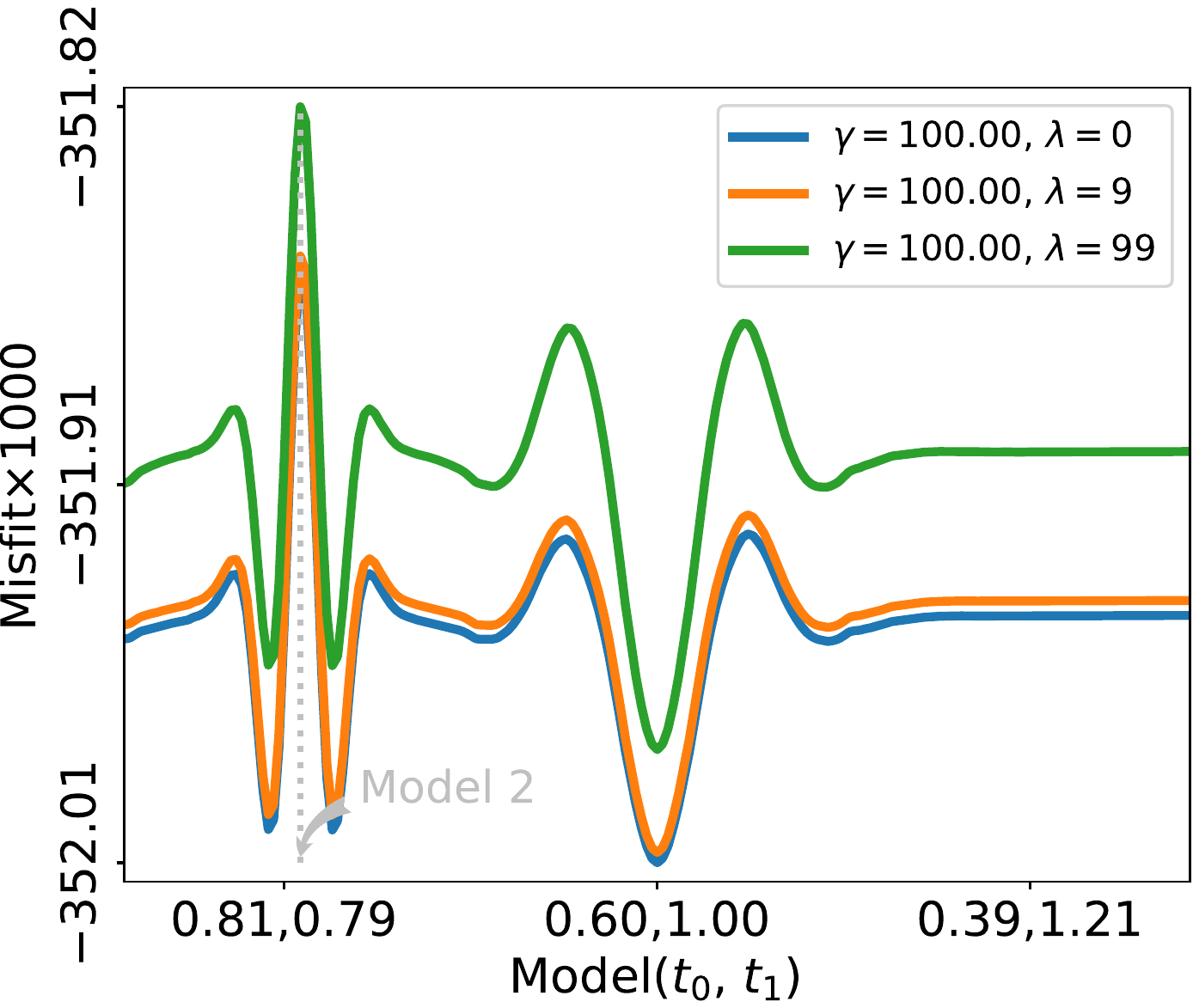}
         \caption{}
         \label{fig:three sin x}
     \end{subfigure}
      \caption{\label{fig:details_dtw}(a-c) the diagonal and (d-f) anti-diagonal details of misfit maps in Figure \ref{fig:dtw_msf}.}
\end{figure}

Figure \ref{fig:dtw_msf} illustrates the normalized misfit measured by the penalized differentiable DTW distance $\varepsilon_{\gamma, \lambda}$, where the value of $\gamma$ keeps the same for every row and the value of $\lambda$ keeps the same for every column. $\gamma$ takes $(0.01,1,100)$ from the top to the bottom and $\lambda$ takes $(0,9,99)$ from the left to the right, respectively. Figure \ref{fig:details_dtw}a-\ref{fig:details_dtw}c and \ref{fig:details_dtw}d-\ref{fig:details_dtw}f show the diagonal and anti-diagonal details of misfit maps from the penalized differentiable DTW. Unlike misfit values in Figure \ref{fig:dtw_msf}, all lines in Figure \ref{fig:details_dtw} represent the un-normalized misfit. We see that the differentiable DTW distance could present negative values because of  the relaxation of the min operator \cite[]{pmlr-v130-blondel21a}. This is also observed in entropy-regularized optimal transport misfit \cite[]{self.Sinkhorn}. From Figure \ref{fig:dtw_msf} and Figure \ref{fig:details_dtw} we conclude: (1) the smoothness of differentiable DTW misfit improves with larger $\gamma$ by comparing different rows in Figure \ref{fig:dtw_msf}; (2) with larger $\gamma$, the differentiable DTW misfit tends to resemble the misfit function based on $\mathcal{L}^2$-norm by comparing Figure \ref{fig:l2_msf} with Figure \ref{fig:dtw_msf}g-\ref{fig:dtw_msf}i, Figure \ref{fig:details_l2}a with Figure \ref{fig:details_dtw}c, or Figure \ref{fig:details_l2}b with Figure \ref{fig:details_dtw}f; (3) for given $\gamma$, the stronger penalization makes the differentiable DTW misfit more identifiable. For example, All green lines in Figure \ref{fig:details_dtw} show a larger range of misfit variation compared to orange and blue lines. 
\begin{figure}[H]
     \centering
     \begin{subfigure}[b]{0.23\textwidth}
         \centering
         \includegraphics[width=\textwidth]{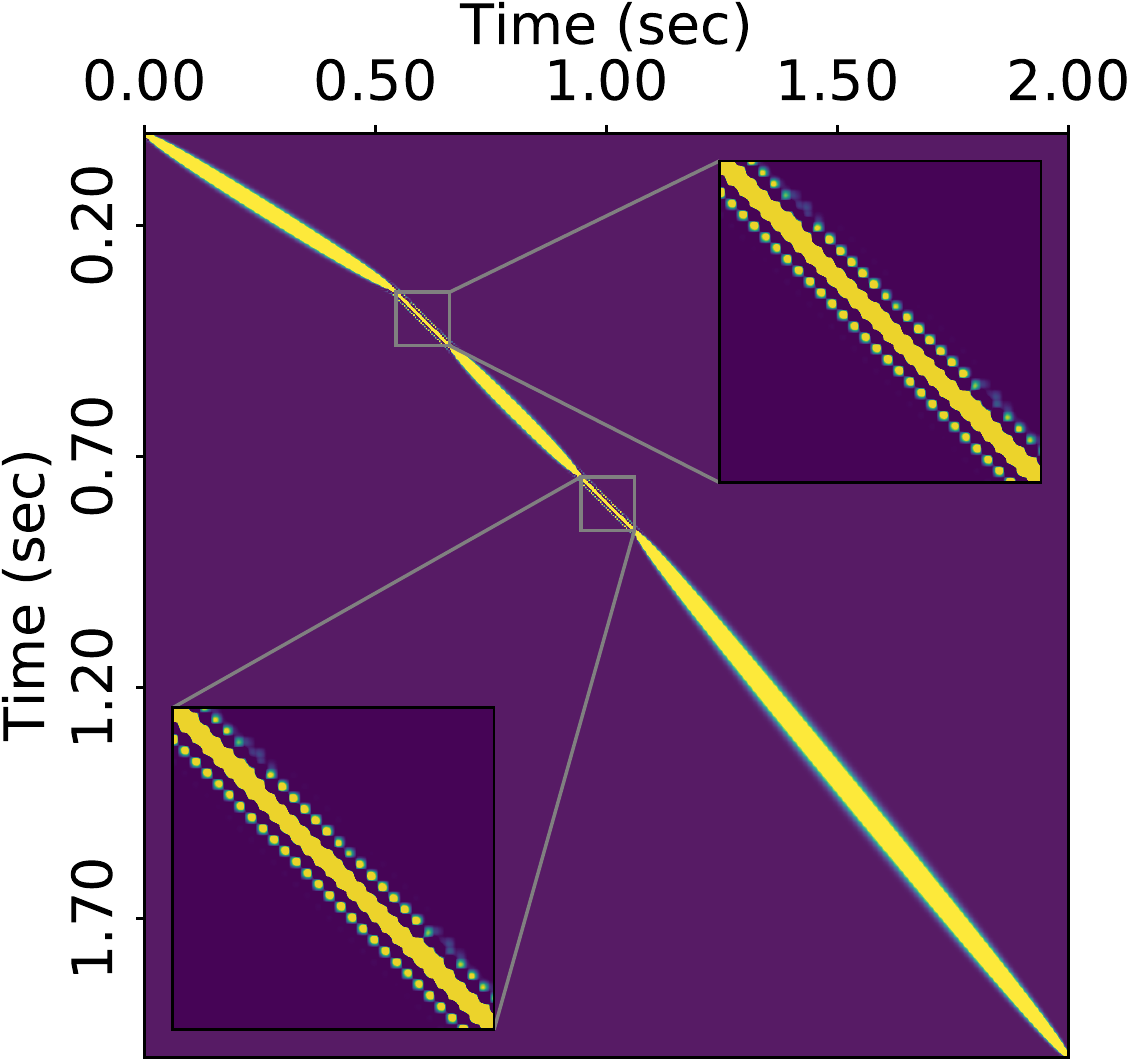}
         \caption{}
         \label{fig:y equals x}
     \end{subfigure}
     \begin{subfigure}[b]{0.23\textwidth}
         \centering
         \includegraphics[width=\textwidth]{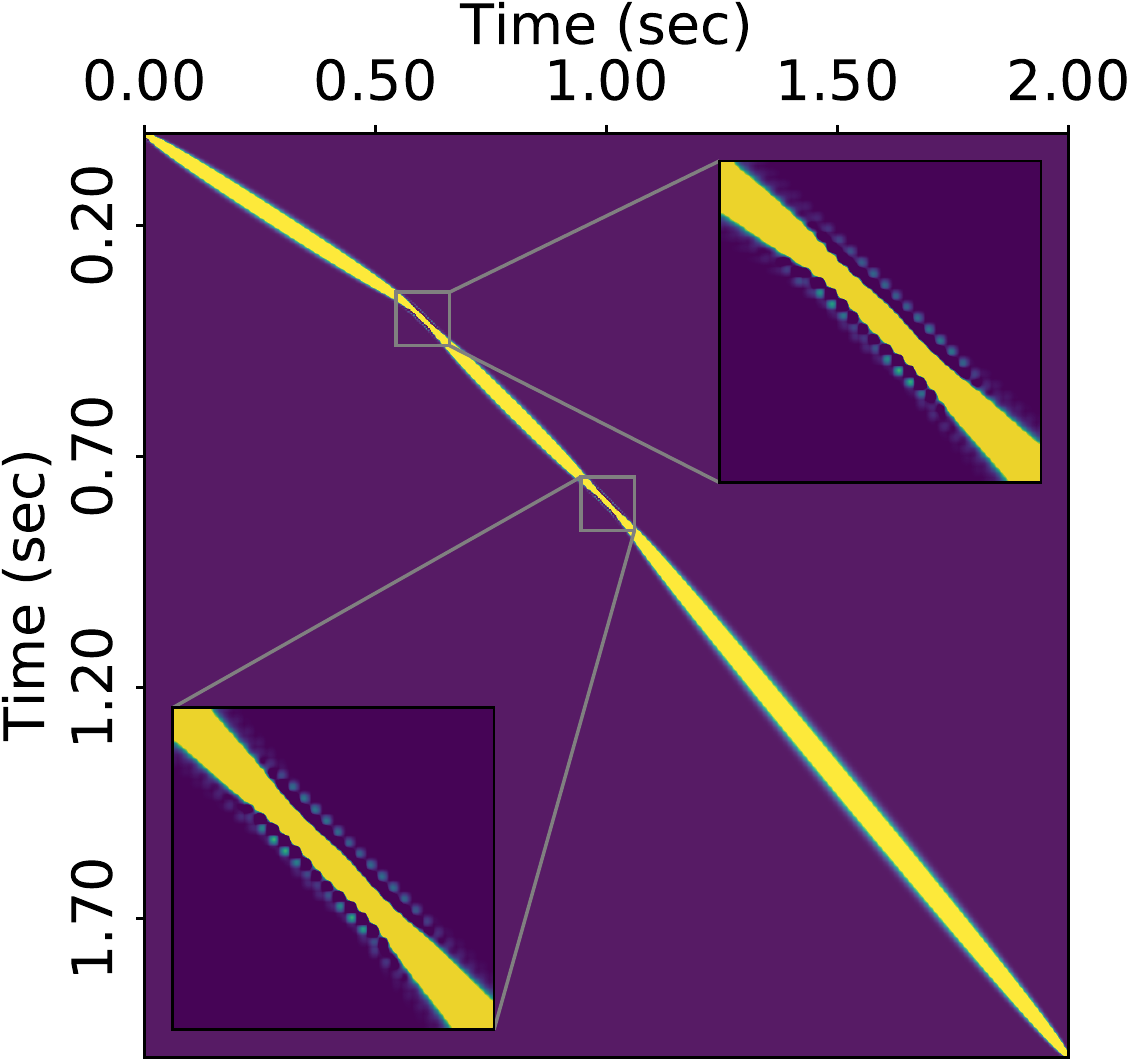}
         \caption{}
         \label{fig:three sin x}
     \end{subfigure}
          \begin{subfigure}[b]{0.23\textwidth}
         \centering
         \includegraphics[width=\textwidth]{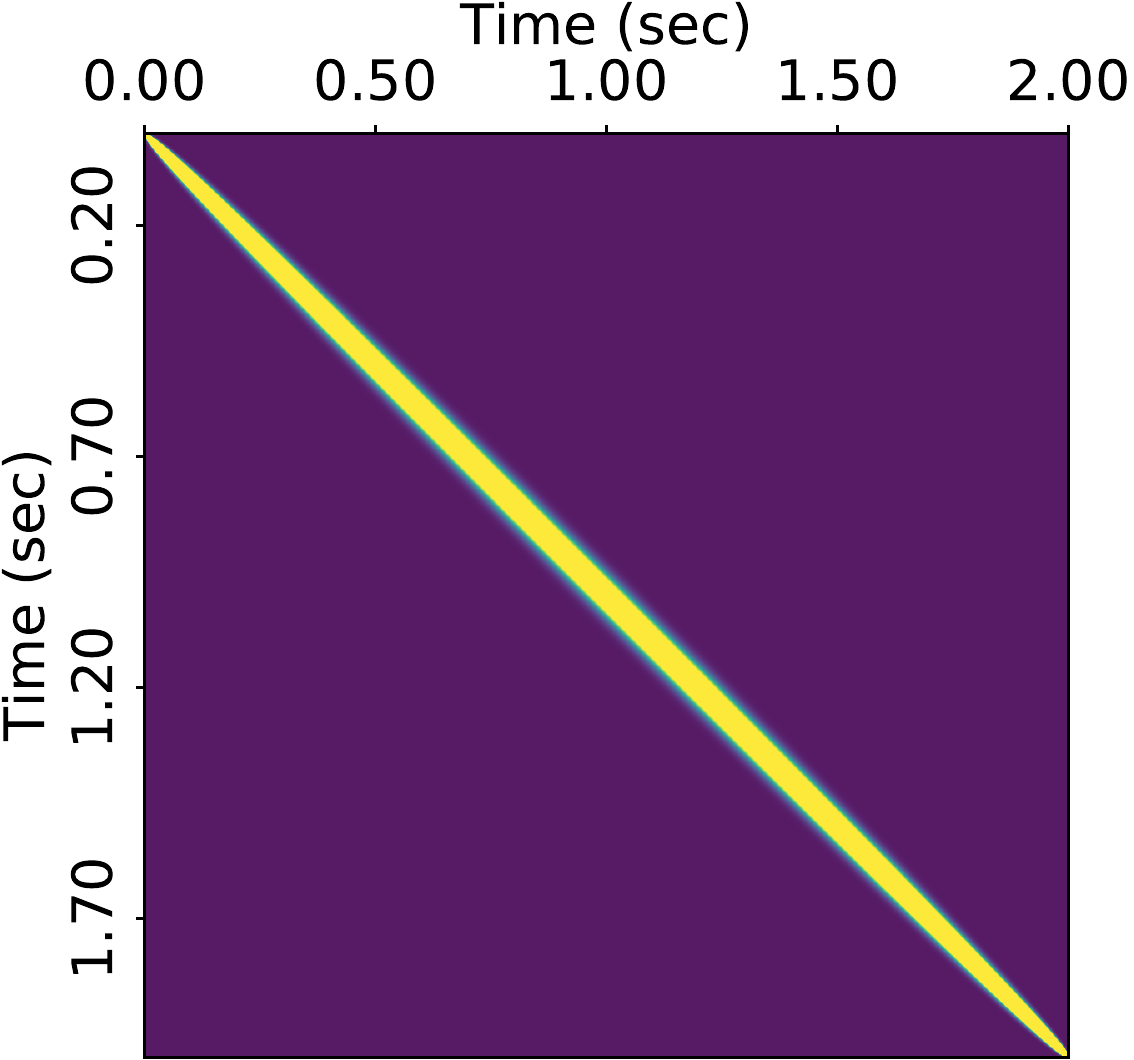}
         \caption{}
         \label{fig:y equals x}
     \end{subfigure}
      \caption{\label{fig:wp_m1} The warping paths for observed and synthetic data in Figure \ref{fig:rickers}a with (a) $\gamma=0.01$, (b) $\gamma=1$, and (c) $\gamma=100$.}
\end{figure}
\begin{figure}[H]
     \centering
     \begin{subfigure}[b]{0.23\textwidth}
         \centering
         \includegraphics[width=\textwidth]{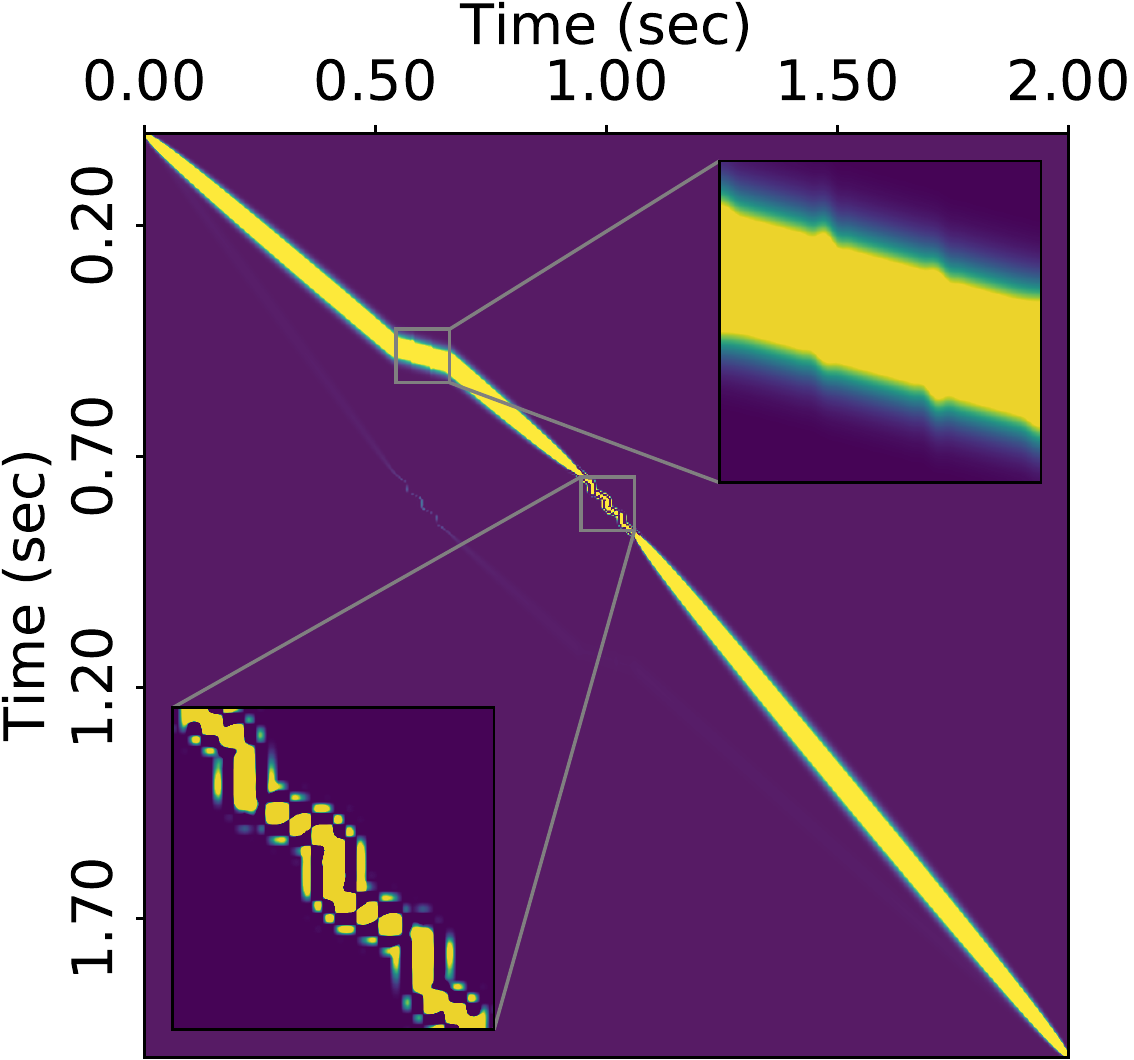}
         \caption{}
         \label{fig:y equals x}
     \end{subfigure}
     \begin{subfigure}[b]{0.23\textwidth}
         \centering
         \includegraphics[width=\textwidth]{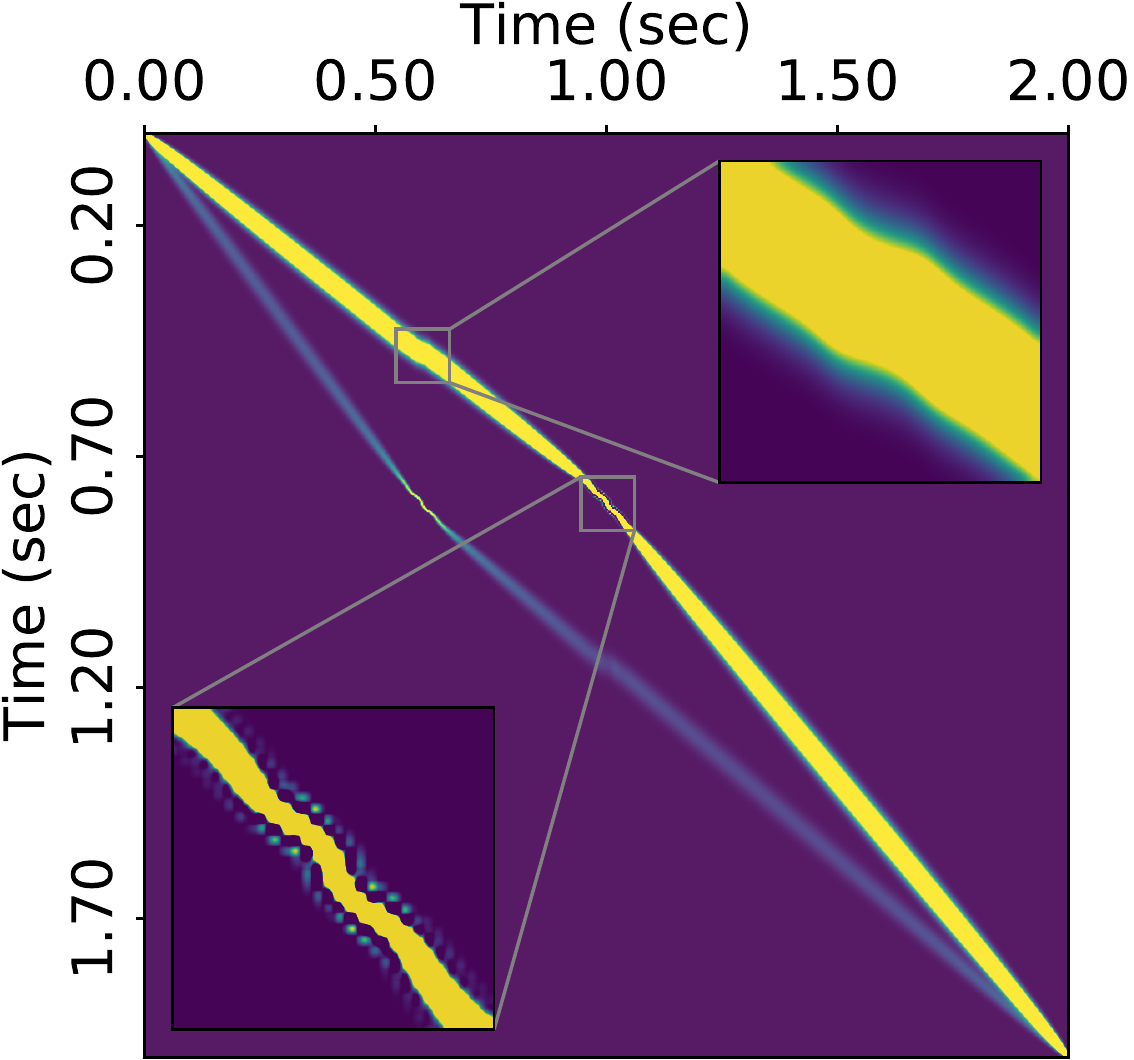}
         \caption{}
         \label{fig:three sin x}
     \end{subfigure}
          \begin{subfigure}[b]{0.23\textwidth}
         \centering
         \includegraphics[width=\textwidth]{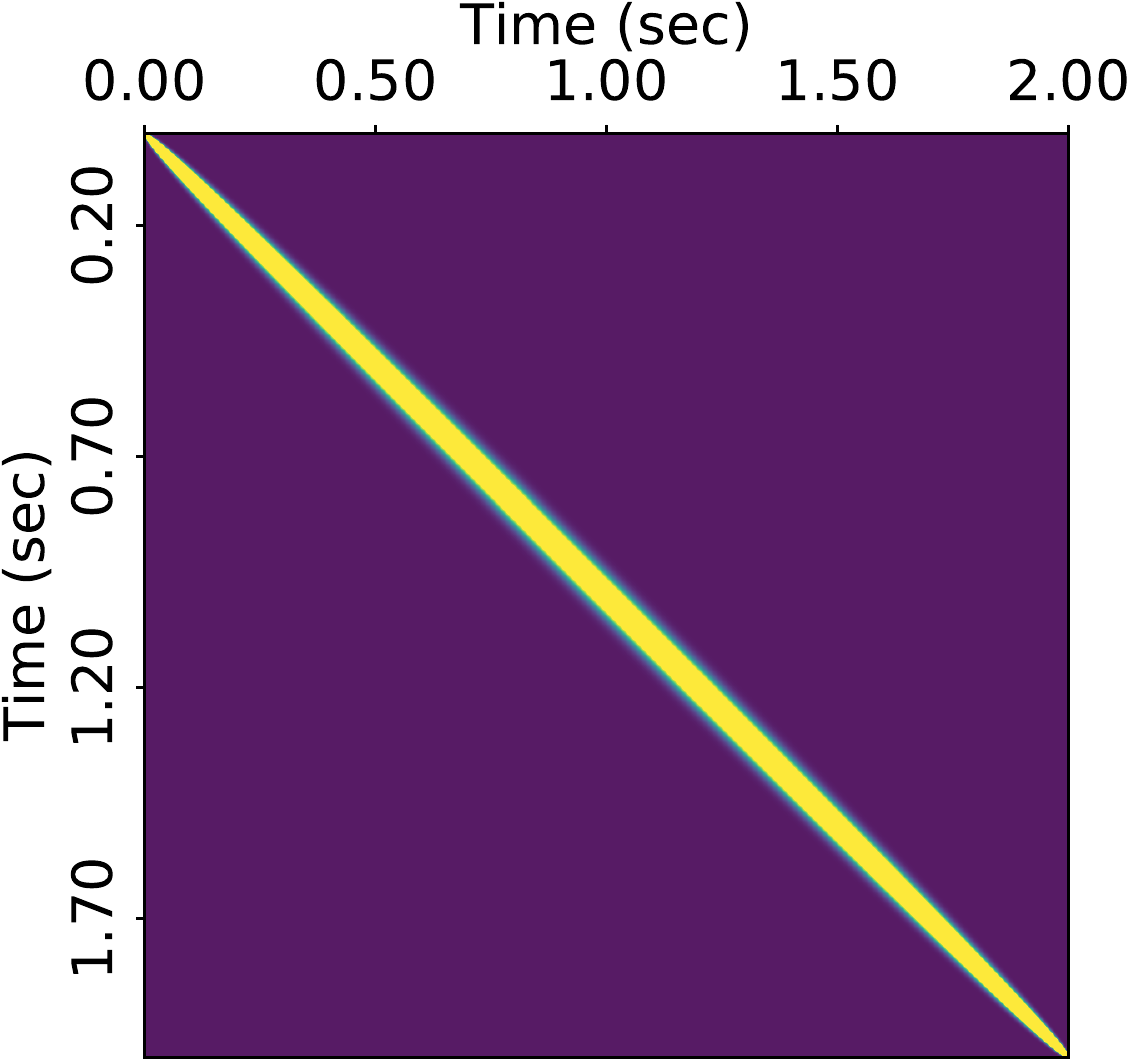}
         \caption{}
         \label{fig:y equals x}
     \end{subfigure}
      \caption{\label{fig:wp_m2} The warping paths for observed and synthetic data in Figure \ref{fig:rickers}b with (a) $\gamma=0.01$, (b) $\gamma=1$, and (c) $\gamma=100$.}
\end{figure}
To further investigate the property of differentiable DTW misfit, we use the observed and synthetic pair in Figure \ref{fig:rickers}a and \ref{fig:rickers}b to generate the warping path, and the results are shown in Figure \ref{fig:wp_m1} and Figure \ref{fig:wp_m2}, respectively. An instant observation from Figure \ref{fig:wp_m1} and Figure \ref{fig:wp_m2} is that the warping path from a larger $\gamma$, for example, $\gamma=100$ in Figure \ref{fig:wp_m1}c and \ref{fig:wp_m2}c makes the differentiable DTW distance approach the weighted $\mathcal{L}^2$-norm. This explains why the differentiable DTW misfit with a larger smoothness parameter $\gamma$ resembles the misfit based on $\mathcal{L}^2$-norm, as we observed above by comparing, for example, Figure \ref{fig:l2_msf} with Figure \ref{fig:dtw_msf}g. The warping path from a smaller gamma, for example, $\gamma=0.01$ (Figure \ref{fig:wp_m1}a and \ref{fig:wp_m2}a) and $\gamma=1$ (Figure \ref{fig:wp_m1}b and \ref{fig:wp_m2}b) will present the transport-like effect. This property makes DTW misfit suitable to be an indicator of traveltime difference. From the zoomed-in plots in Figure \ref{fig:wp_m1} and \ref{fig:wp_m2}, we see a smaller $\gamma$ will result in a sharper or less smooth warping path.

\begin{wrapfigure}{R}{9cm}
\centering
\includegraphics[width=.5\textwidth]{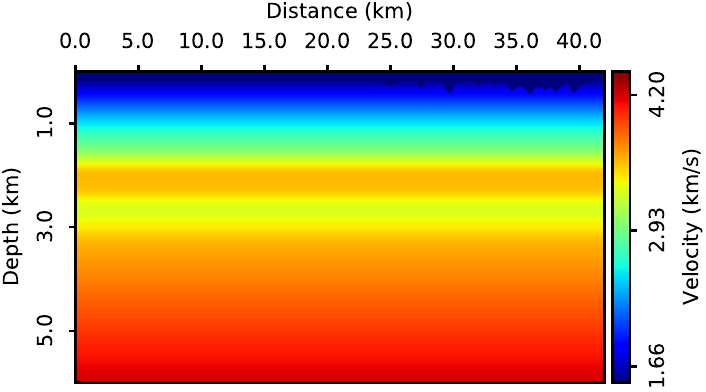}
    \caption{\label{fig:init_gom14} Initial starting model released with the Chevron blind test datasets.}
\end{wrapfigure}

\begin{figure}[H]
     \centering
     \begin{subfigure}[b]{0.43\textwidth}
         \centering
         \includegraphics[width=\textwidth]{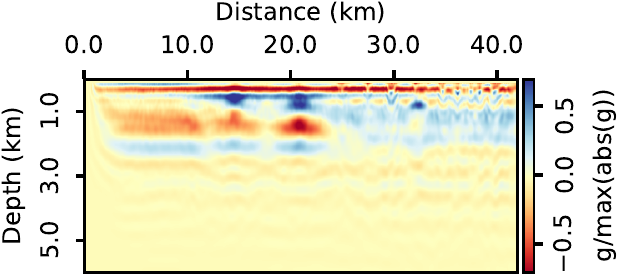}
         \caption{}
         \label{fig:y equals x}
     \end{subfigure}
     \begin{subfigure}[b]{0.43\textwidth}
         \centering
         \includegraphics[width=\textwidth]{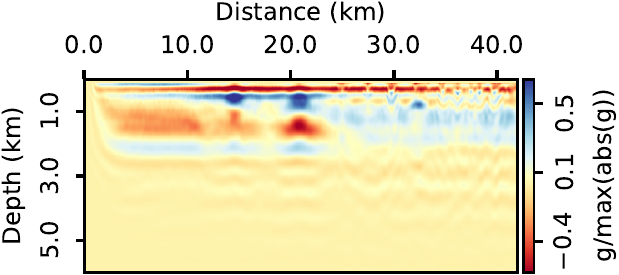}
         \caption{}
         \label{fig:three sin x}
     \end{subfigure}
          \begin{subfigure}[b]{0.43\textwidth}
         \centering
         \includegraphics[width=\textwidth]{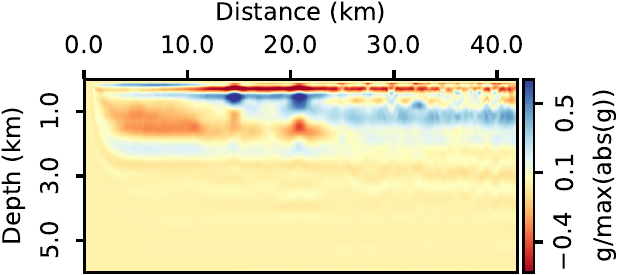}
         \caption{}
         \label{fig:y equals x}
     \end{subfigure}
     \begin{subfigure}[b]{0.43\textwidth}
         \centering
         \includegraphics[width=\textwidth]{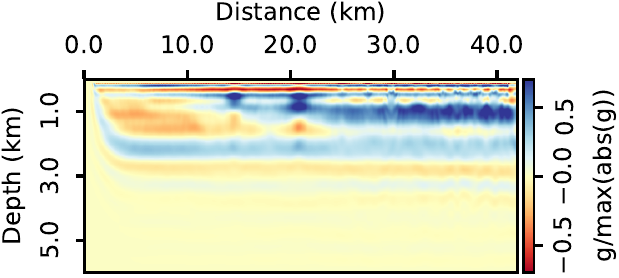}
         \caption{}
         \label{fig:y equals x}
     \end{subfigure}
      \caption{\label{fig:grad0} The gradient of penalized differentiable DTW distance from $\lambda=0$ and $\gamma={[0.1, 1, 10, 100]}$ for (a), (b), (c) and (d). The same order is used for Figures \ref{fig:grad9} and \ref{fig:grad99}. 
We normalize every gradient in ${[-1, 1]}$ by doing g$\leftarrow$g/abs(g).max(), also for those in Figures \ref{fig:grad9} and \ref{fig:grad99}.}
\end{figure}

Next, we use Chevron blind test datasets to illustrate (a) how the smooth parameter $\gamma$ and penalization parameter $\lambda$ affect the adjoint source and therefore the gradient, (b) the pitfall of the conventional non-differentiable DTW misfit function, and (c) demonstrate that the differentiable one can remedy this issue. 
We start with the initial model in Figure \ref{fig:init_gom14} and data in a frequency range {$\SI{3.0}{\hertz}$-$\SI{3.5}{\hertz}$}. We first invert the early arrivals only because in this frequency range the reflection data are dominated by noise. Figure \ref{fig:adj0}a-\ref{fig:adj0}d shows the adjoint sources from different smoothness parameters $\gamma={[0.1,1,10,100]}$ in order and penalization parameter $\lambda=0$. An instant observation from Figure \ref{fig:adj0} is that a smaller $\gamma$ such as $\gamma=0.1$ (Figure \ref{fig:adj0}a) and $\gamma=1$ (Figure \ref{fig:adj0}b) generates stronger discontinuities in the adjoint source while a larger $\gamma$ such as $\gamma=10$ (Figure \ref{fig:adj0}c) and $\gamma=100$ (Figure \ref{fig:adj0}d) will eliminate these abrupt changes. Considering that the differentiable DTW distance approaches the non-differentiable one when $\gamma$ approaches zero, we conclude that the nondifferentible DTW distance will generate more severe abrupt changes in the adjoint source than the one in Figure \ref{fig:adj0}a ($\gamma=0.1$ and $\lambda=0$). We claim that this property makes the conventional non-differentiable DTW distance less suitable to mitigate the local minima issue. The results of gradient from adjoint sources in Figure \ref{fig:adj0} are illustrated in Figure \ref{fig:grad0}, where we see that 
the adjoint source from a smaller $\gamma$ results in a gradient with higher wavenumbers. However, we prefer lower wavenumbers in perturbation estimated from lower frequency data.

\begin{figure}[H]
     \centering
     \begin{subfigure}[b]{0.23\textwidth}
         \centering
         \includegraphics[width=\textwidth]{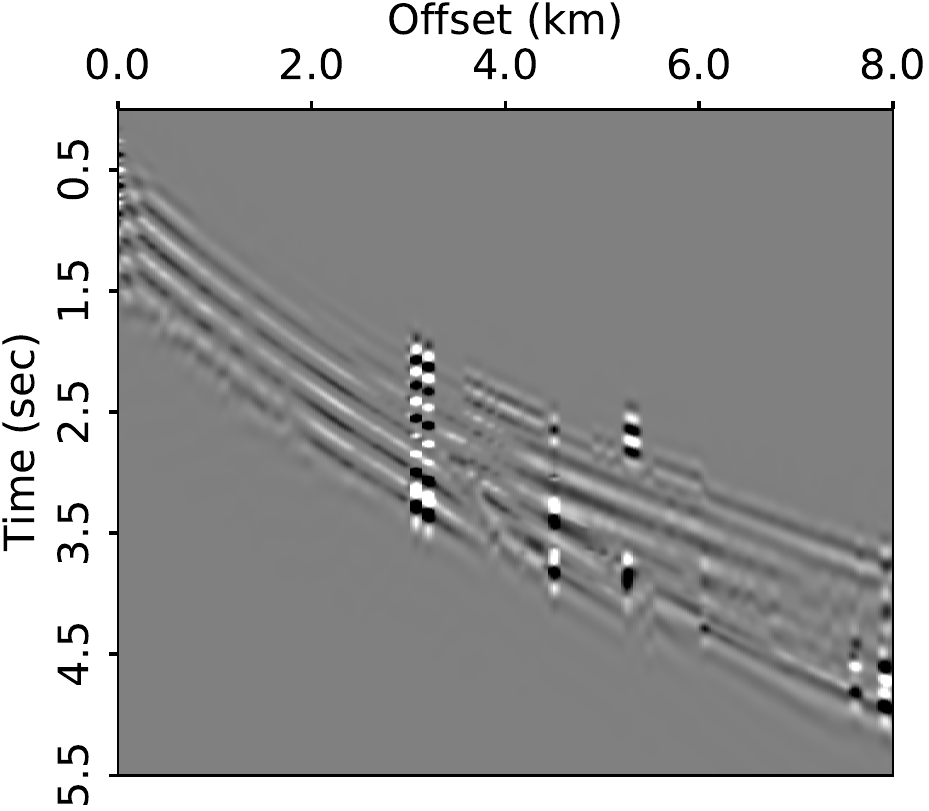}
         \caption{}
         \label{fig:y equals x}
     \end{subfigure}
     \begin{subfigure}[b]{0.23\textwidth}
         \centering
         \includegraphics[width=\textwidth]{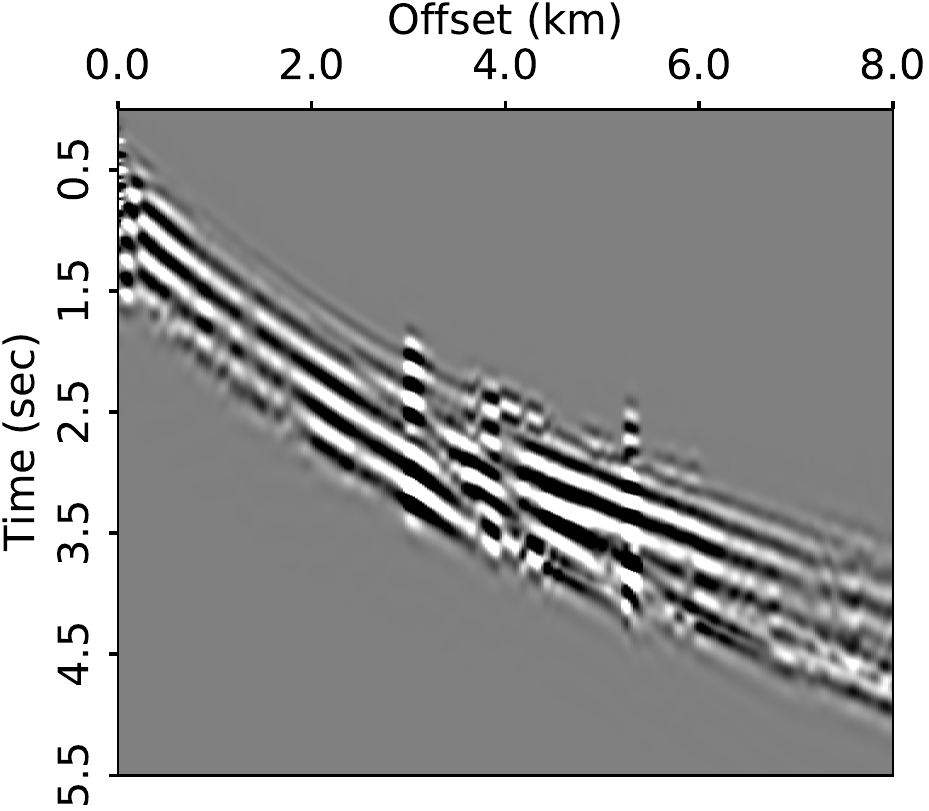}
         \caption{}
         \label{fig:three sin x}
     \end{subfigure}
          \begin{subfigure}[b]{0.23\textwidth}
         \centering
         \includegraphics[width=\textwidth]{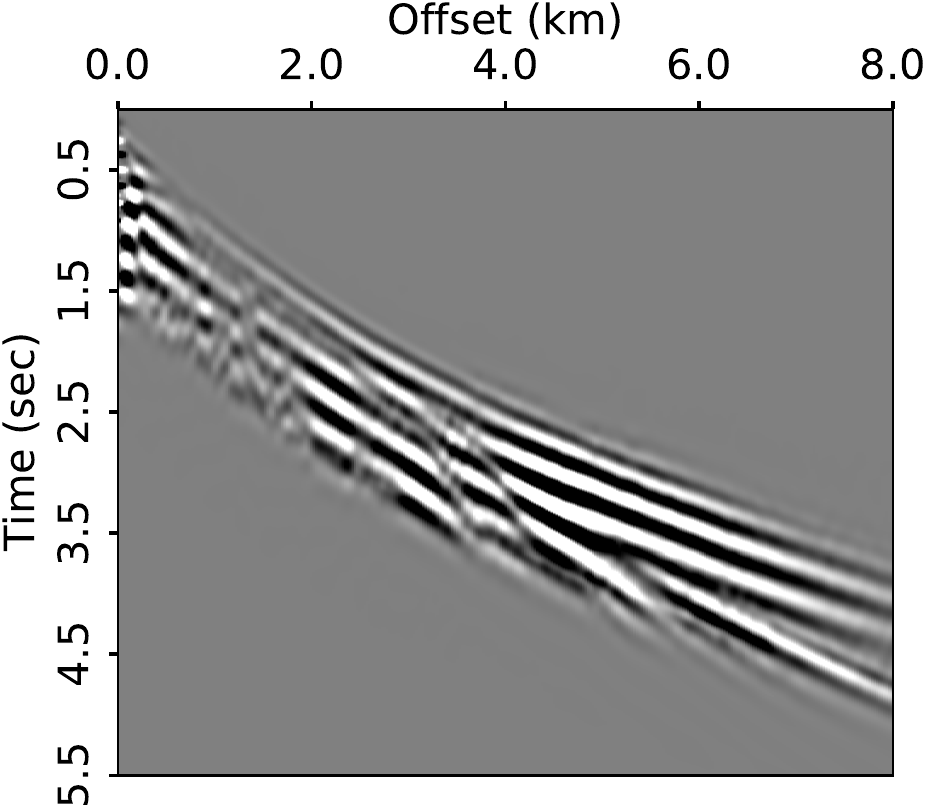}
         \caption{}
         \label{fig:y equals x}
     \end{subfigure}
     \begin{subfigure}[b]{0.23\textwidth}
         \centering
         \includegraphics[width=\textwidth]{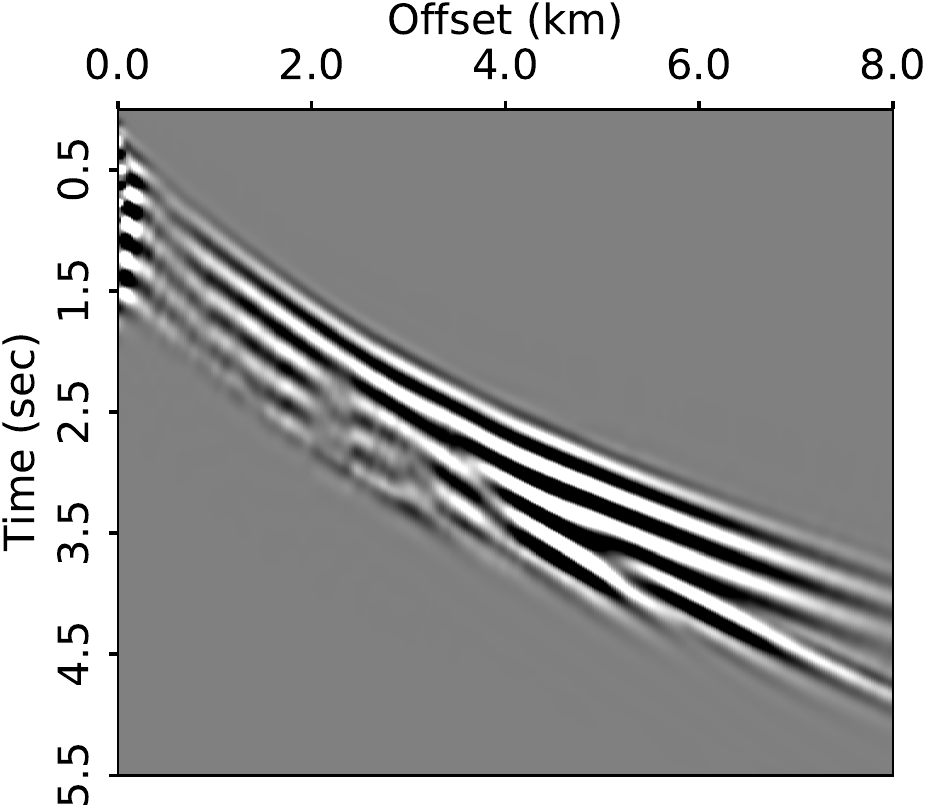}
         \caption{}
         \label{fig:y equals x}
     \end{subfigure}
      \caption{\label{fig:adj9} The adjoint source of penalized differentiable DTW distance from $\lambda=9$ and $\gamma={[0.1, 1, 10, 100]}$. }
\end{figure}

\begin{figure}[H]
     \centering
     \begin{subfigure}[b]{0.43\textwidth}
         \centering
         \includegraphics[width=\textwidth]{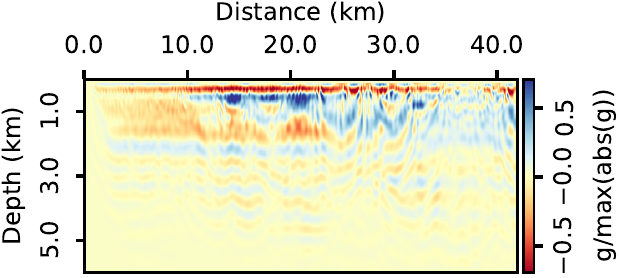}
         \caption{}
         \label{fig:y equals x}
     \end{subfigure}
     \begin{subfigure}[b]{0.43\textwidth}
         \centering
         \includegraphics[width=\textwidth]{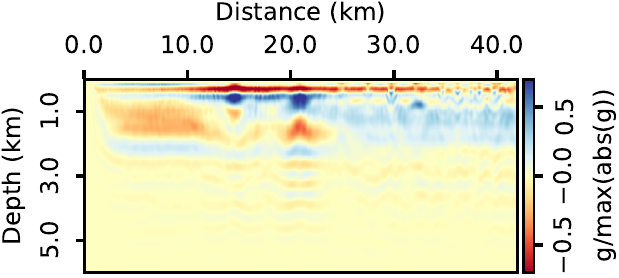}
         \caption{}
         \label{fig:three sin x}
     \end{subfigure}
          \begin{subfigure}[b]{0.43\textwidth}
         \centering
         \includegraphics[width=\textwidth]{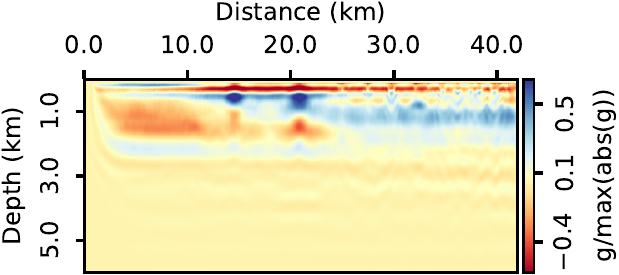}
         \caption{}
         \label{fig:y equals x}
     \end{subfigure}
     \begin{subfigure}[b]{0.43\textwidth}
         \centering
         \includegraphics[width=\textwidth]{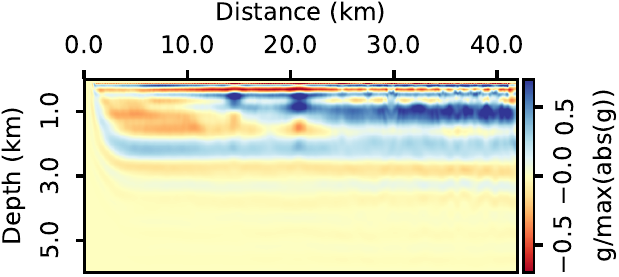}
         \caption{}
         \label{fig:y equals x}
     \end{subfigure}
      \caption{\label{fig:grad9} The gradient of penalized differentiable DTW distance from $\lambda=9$ and $\gamma={[0.1, 1, 10, 100]}$.}
\end{figure}

\begin{figure}[H]
     \centering
     \begin{subfigure}[b]{0.23\textwidth}
         \centering
         \includegraphics[width=\textwidth]{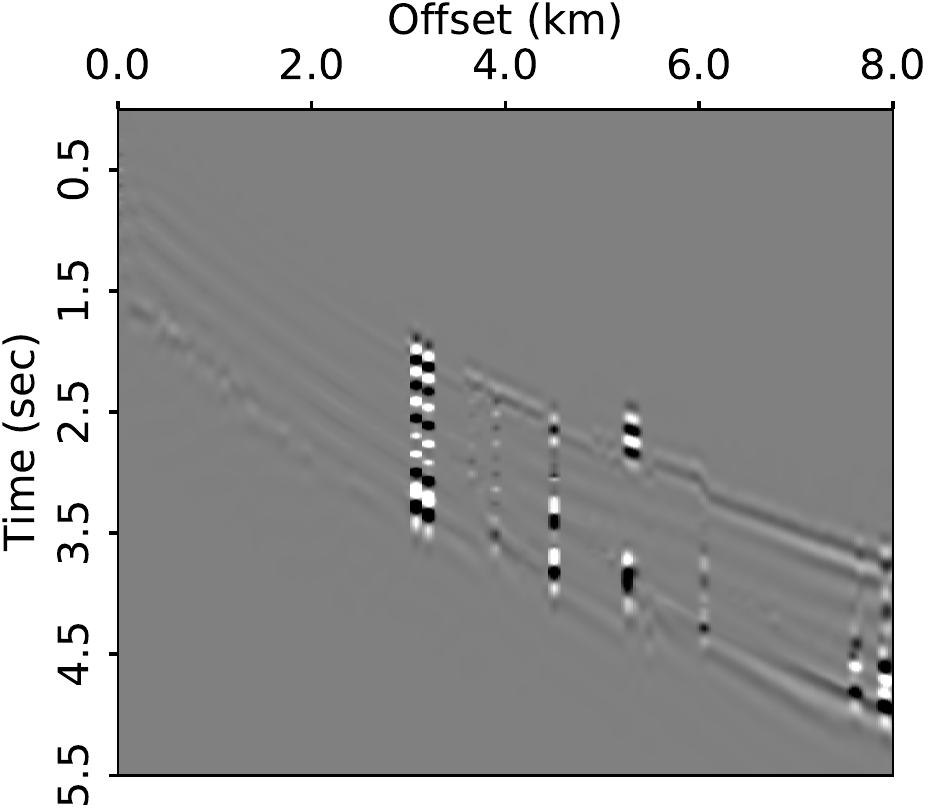}
         \caption{}
         \label{fig:y equals x}
     \end{subfigure}
     \begin{subfigure}[b]{0.23\textwidth}
         \centering
         \includegraphics[width=\textwidth]{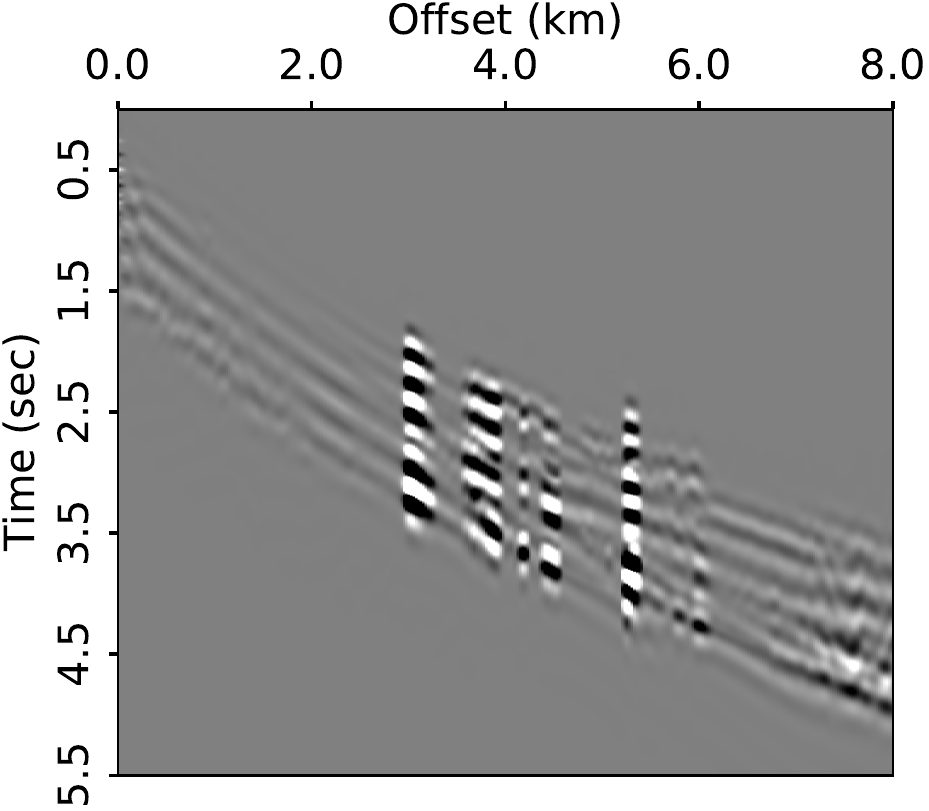}
         \caption{}
         \label{fig:three sin x}
     \end{subfigure}
          \begin{subfigure}[b]{0.23\textwidth}
         \centering
         \includegraphics[width=\textwidth]{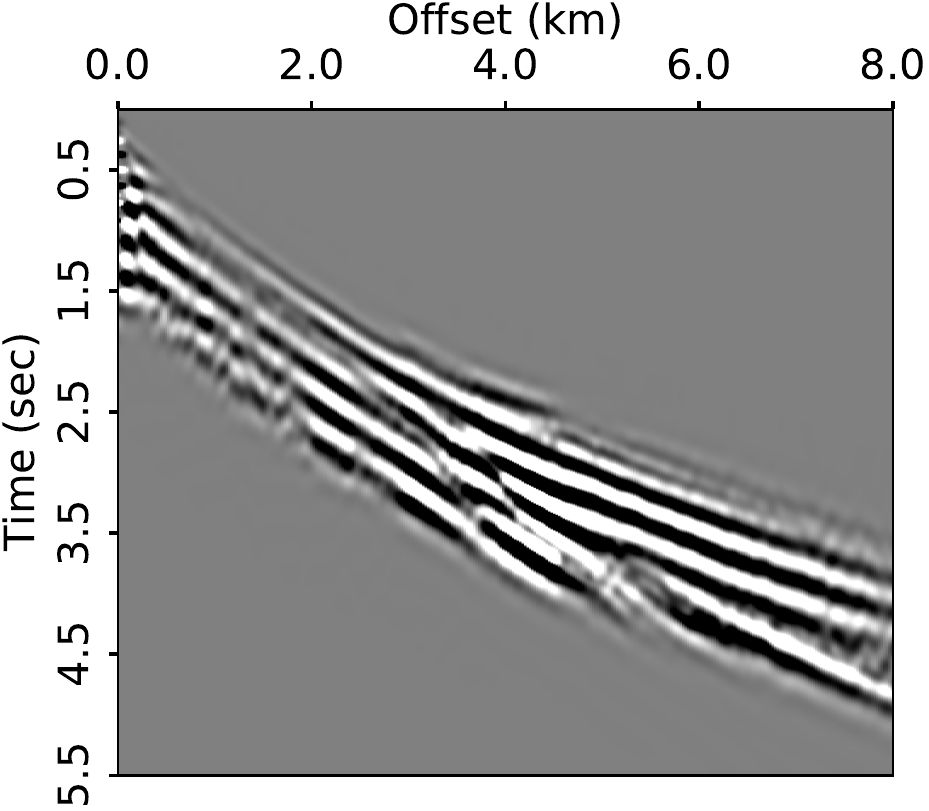}
         \caption{}
         \label{fig:y equals x}
     \end{subfigure}
     \begin{subfigure}[b]{0.23\textwidth}
         \centering
         \includegraphics[width=\textwidth]{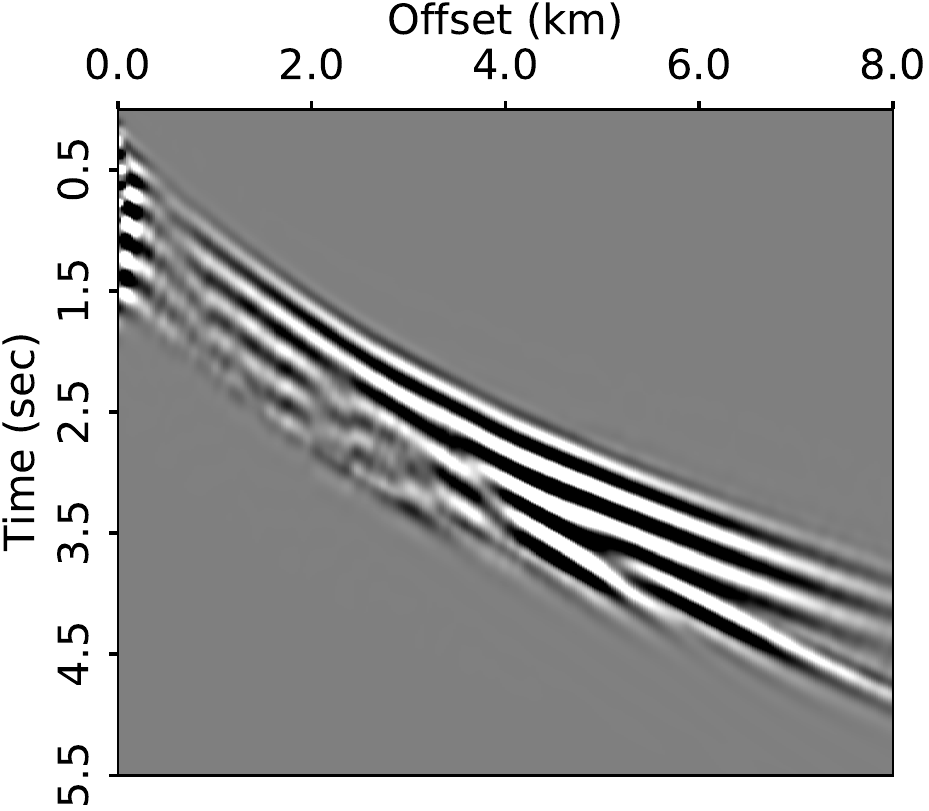}
         \caption{}
         \label{fig:y equals x}
     \end{subfigure}
      \caption{\label{fig:adj0}\label{fig:adj99} The adjoint source of penalized differentiable DTW distance from $\lambda=99$ and $\gamma={[0.1, 1, 10, 100]}$.}
\end{figure}

\begin{figure}[H]
     \centering
     \begin{subfigure}[b]{0.43\textwidth}
         \centering
         \includegraphics[width=\textwidth]{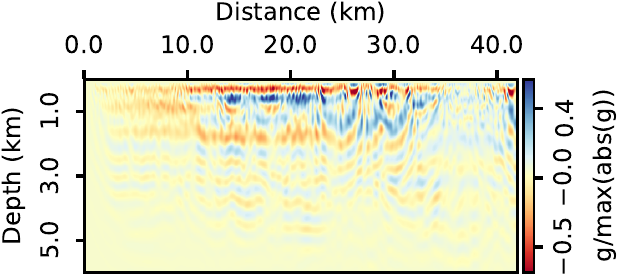}
         \caption{}
         \label{fig:y equals x}
     \end{subfigure}
     \begin{subfigure}[b]{0.43\textwidth}
         \centering
         \includegraphics[width=\textwidth]{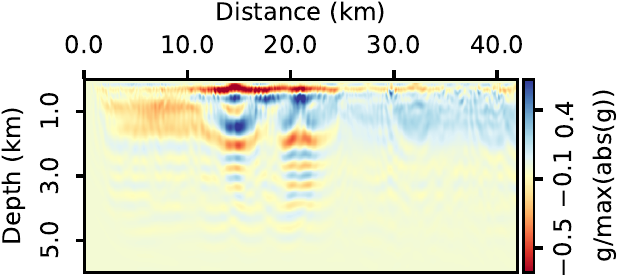}
         \caption{}
         \label{fig:three sin x}
     \end{subfigure}
          \begin{subfigure}[b]{0.43\textwidth}
         \centering
         \includegraphics[width=\textwidth]{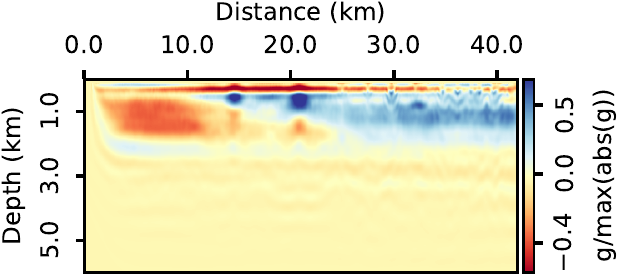}
         \caption{}
         \label{fig:y equals x}
     \end{subfigure}
     \begin{subfigure}[b]{0.43\textwidth}
         \centering
         \includegraphics[width=\textwidth]{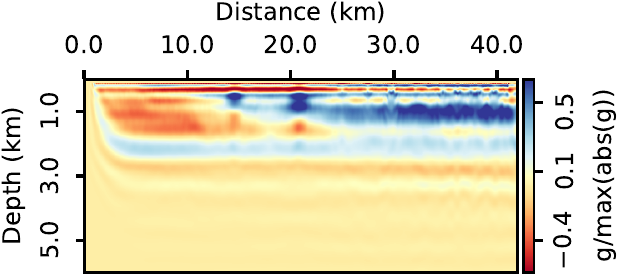}
         \caption{}
         \label{fig:y equals x}
     \end{subfigure}
      \caption{\label{fig:grad99} The gradient of penalized differentiable DTW distance with $\lambda=99$ and $\gamma={[0.1, 1, 10, 100]}$. }
\end{figure}

Figure \ref{fig:adj9}a-\ref{fig:adj9}d shows the adjoint sources from smoothness parameters $\gamma={[0.1,1,10,100]}$ in order, but penalization parameter $\lambda=9$. Similarly, we can conclude that a larger $\gamma$ results in a smoother adjoint source. Figure \ref{fig:grad9} shows the gradient results from the adjoint sources in Figure \ref{fig:adj9} in order, where we see the abrupt changes in adjoint source yield the implausible high wavenumber noise in gradient as demonstrated in Figure \ref{fig:grad9}a-\ref{fig:grad9}c. Figure \ref{fig:adj99} and Figure \ref{fig:grad99} show the adjoint source and the corresponding gradient with the smoothness parameter $\gamma={[0.1,1,10,100]}$ and the penalization parameter $\lambda=99$. Again, we see a larger $\gamma$ can reduce the discontinuity in adjoint source for a given penalization parameter $\lambda$, and Figure \ref{fig:grad99} also reflects this improvement. 

To analyze the effect of the penalization parameter $\lambda$ on the adjoint source, we compare the plots with the same $\gamma$ but different $\lambda$, for example, Figure \ref{fig:adj0}a, \ref{fig:adj9}a, and \ref{fig:adj99}a where $\gamma=0.1$ and $\lambda={[0,9,99]}$. We conclude that a larger penalization parameter $\lambda$ introduces a stronger discontinuity from this comparison for some small $\gamma$. We can also see this phenomena by comparing Figure \ref{fig:adj0}b, \ref{fig:adj9}b, and \ref{fig:adj99}b, where $\gamma=1$ and $\lambda={[0,9,99]}$. For some larger values of $\gamma$, for example $\gamma=10$ as in Figure \ref{fig:adj0}c, \ref{fig:adj9}c, and \ref{fig:adj99}c and $\gamma=100$ as in Figure \ref{fig:adj0}d, \ref{fig:adj9}d, and \ref{fig:adj99}d, the increase in the penalization parameter $\lambda$ does not bring severe abrupt changes any more.

Observations about how the smoothness parameter $\gamma$ and penalization parameter $\lambda$ shapes the adjoint source can be corroborated with mathematical arguments \cite[]{{Cuturi2017SoftDTWAD,MenschBlondel2018}}. The DTW distance $\varepsilon_\gamma$ and its gradient are $1/\gamma$- and $2/\gamma$-Lipschitz continuous, respectively. An $\alpha$-Lipschitz continuous function means its change rate is bounded by $\alpha$. Therefore, a smaller $\gamma$, for example, $\gamma=0.01$ which makes $\min^\gamma$ equal to $\min$ almost for every subproblem in this example, allows larger change rates. This is why we see the sharp changes in adjoint sources as shown in Figures \ref{fig:adj0}a, \ref{fig:adj0}b, \ref{fig:adj9}a,\ref{fig:adj9}b, \ref{fig:adj99}a, and \ref{fig:adj99}b. By comparing Figure \ref{fig:adj99}b with Figures \ref{fig:adj9}b and Figure \ref{fig:adj0}b, we see stronger penalization results in severer sharp changes because the adjoint source induced by the penalization term involves the second-order derivative, that admits a larger Lipschitz constant than $2/\gamma$. Besides, a smaller $\gamma$ results in a larger difference between adjoint sources calculated by the definition and the equation \ref{eq:adj}. A larger $\gamma$ can significantly reduce this discrepancy.
\begin{figure}[H]
     \centering
     \begin{subfigure}[b]{0.23\textwidth}
         \centering
         \includegraphics[width=\textwidth]{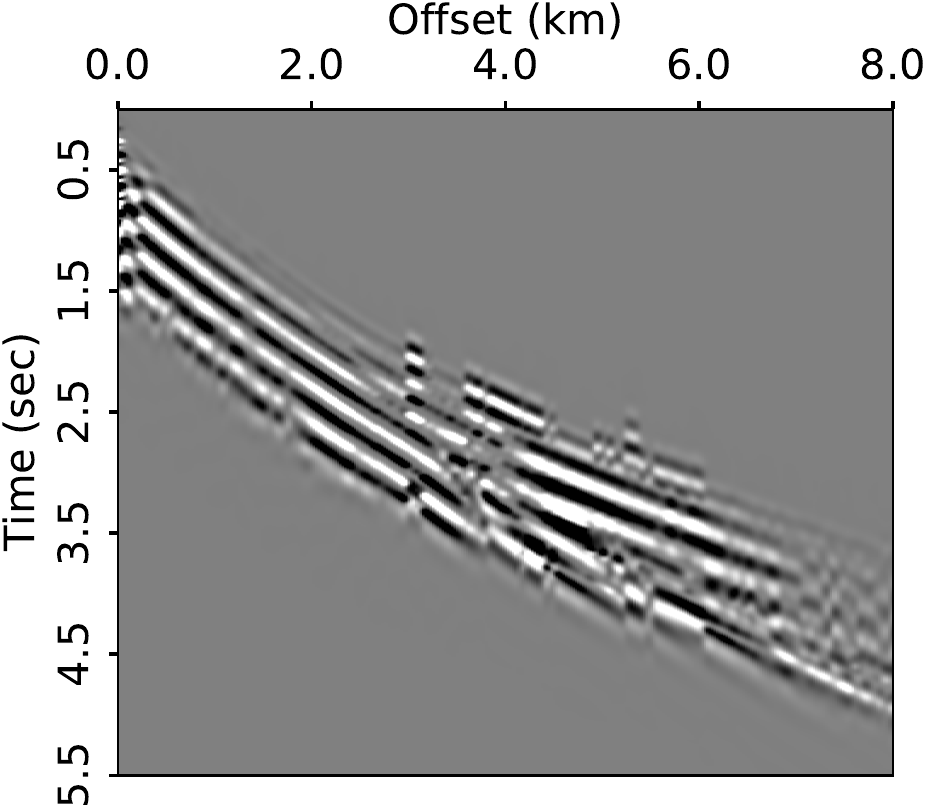}
         \caption{}
         \label{fig:y equals x}
     \end{subfigure}
     \begin{subfigure}[b]{0.23\textwidth}
         \centering
         \includegraphics[width=\textwidth]{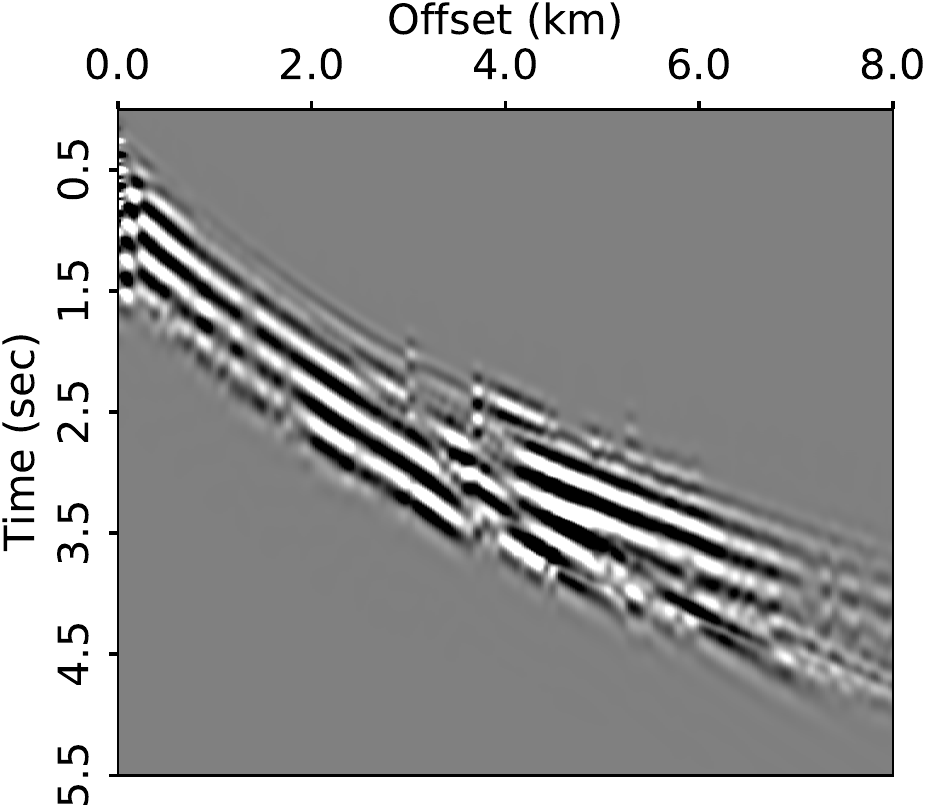}
         \caption{}
         \label{fig:three sin x}
     \end{subfigure}
          \begin{subfigure}[b]{0.23\textwidth}
         \centering
         \includegraphics[width=\textwidth]{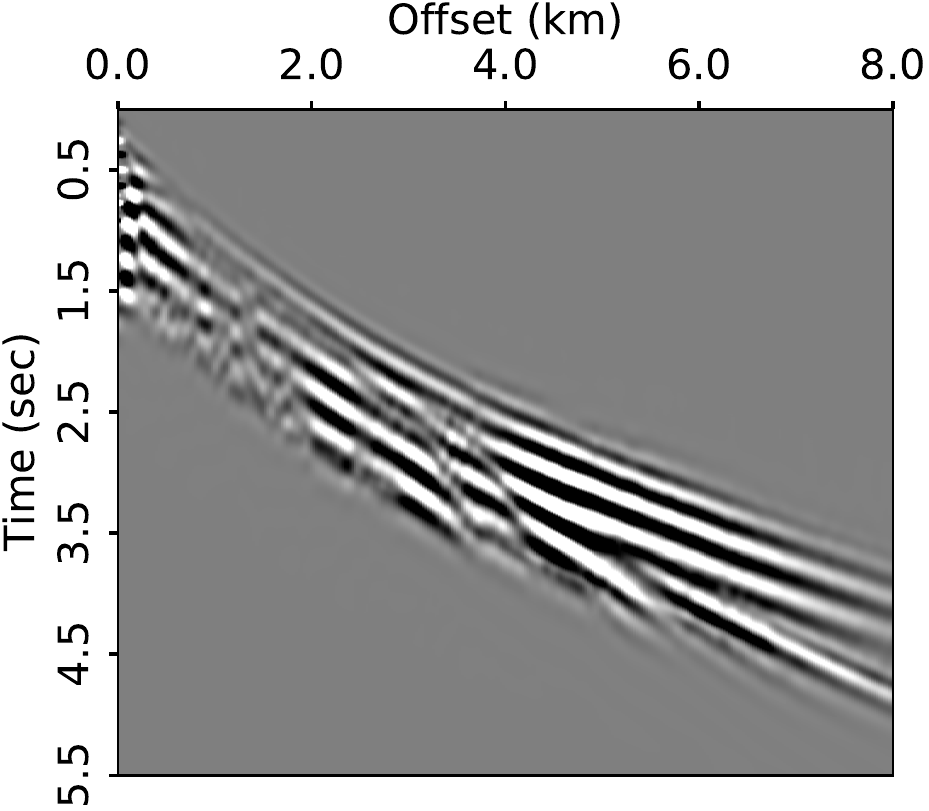}
         \caption{}
         \label{fig:y equals x}
     \end{subfigure}
     \begin{subfigure}[b]{0.23\textwidth}
         \centering
         \includegraphics[width=\textwidth]{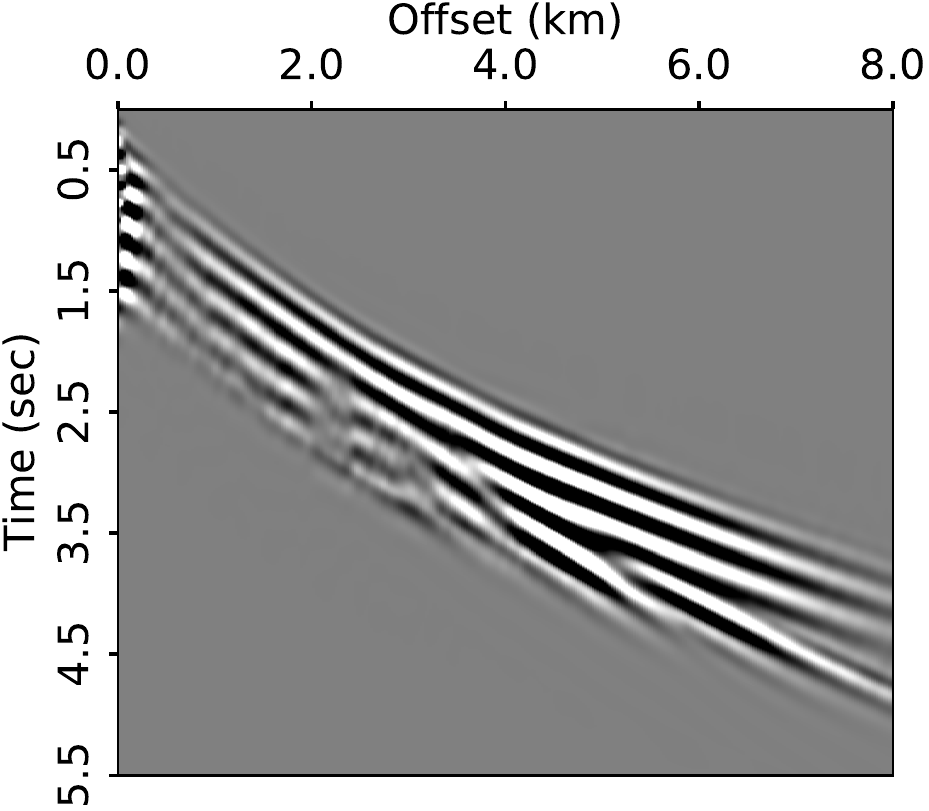}
         \caption{}
         \label{fig:y equals x}
     \end{subfigure}
      \caption{\label{fig:adj0} The adjoint sources of penalized differentiable DTW distance from $\lambda=0$ and $\gamma={[0.1, 1, 10, 100]}$ for (a), (b), (c) and (d). The same order is used for Figures \ref{fig:adj9} and \ref{fig:adj99}.}
\end{figure}

\begin{figure}[H]
     \centering
     \begin{subfigure}[b]{0.43\textwidth}
         \centering
         \includegraphics[width=\textwidth]{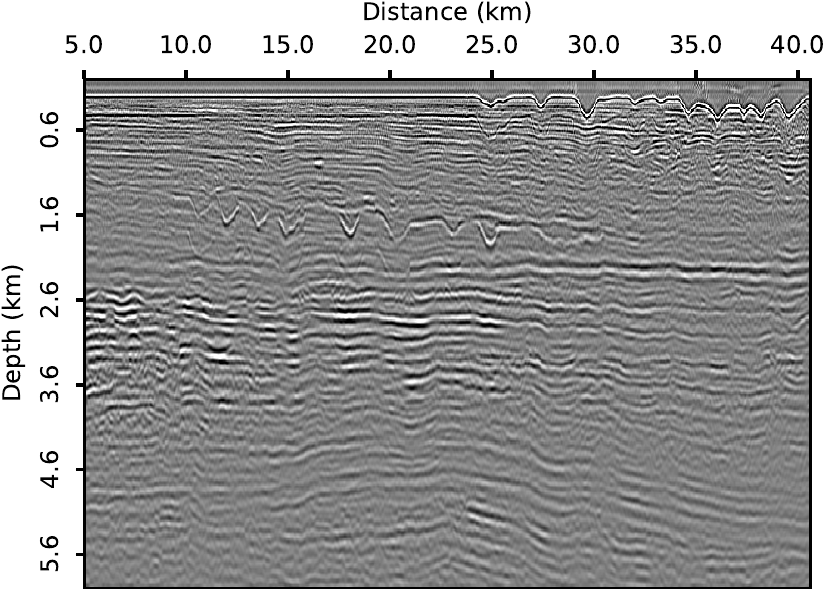}
         \caption{}
         \label{fig:y equals x}
     \end{subfigure}
     \begin{subfigure}[b]{0.43\textwidth}
         \centering
         \includegraphics[width=\textwidth]{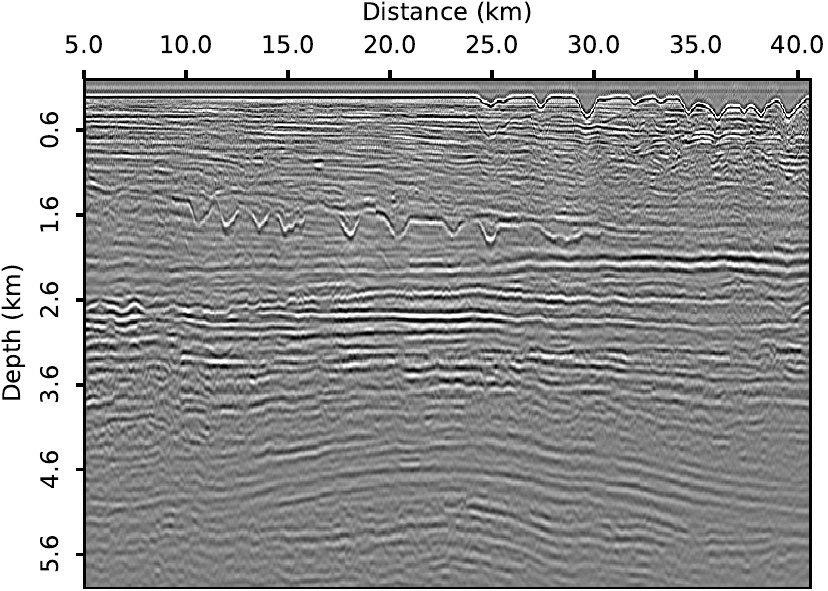}
         \caption{}
         \label{fig:three sin x}
     \end{subfigure}
      \caption{\label{fig:rtm02} The migration result (a) from the initial and (b) from the final inverted model.}
\end{figure}

\begin{figure}[H]
     \centering
     \begin{subfigure}[b]{0.43\textwidth}
         \centering
         \includegraphics[width=\textwidth]{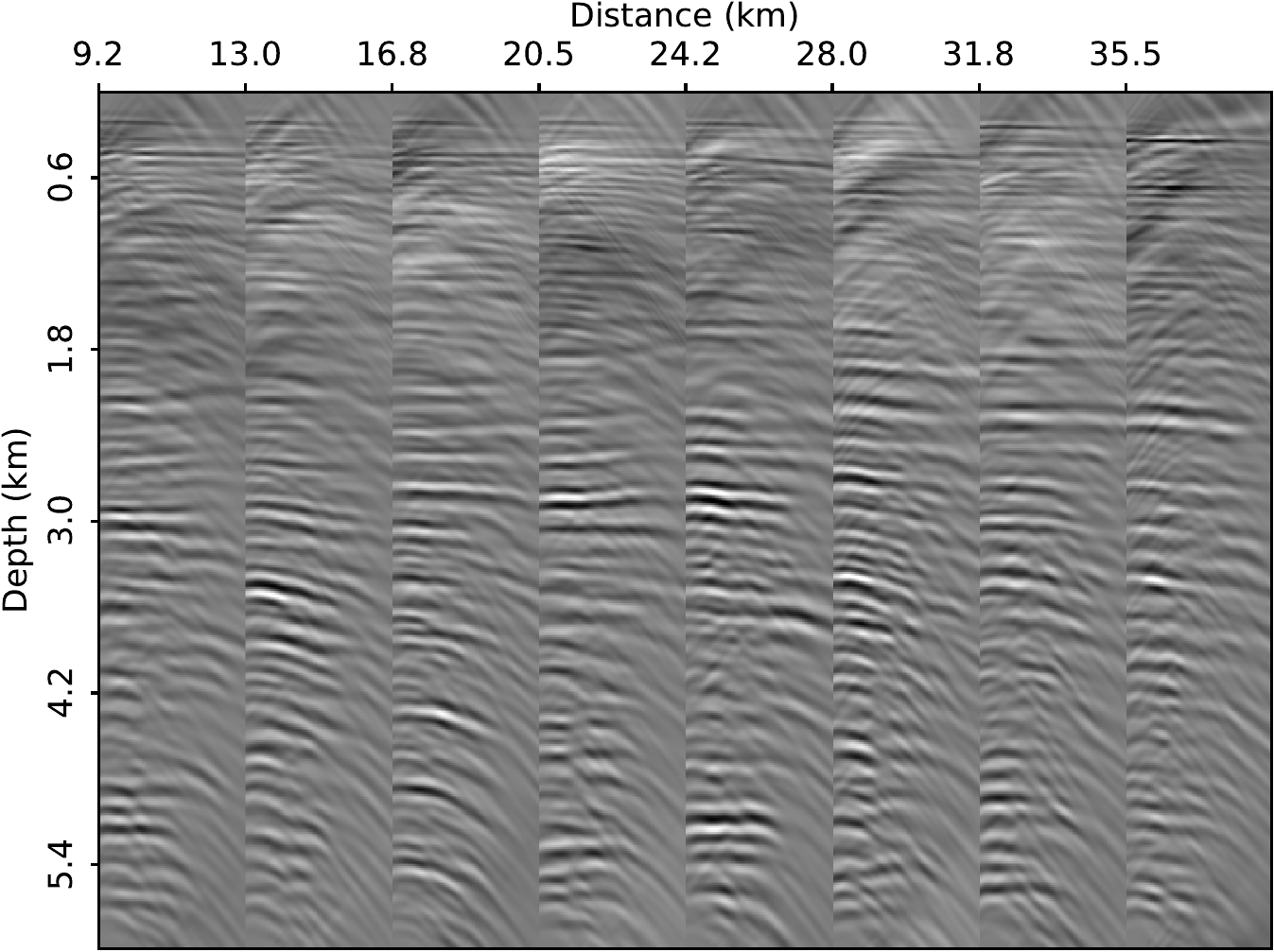}
         \caption{}
         \label{fig:y equals x}
     \end{subfigure}
     \begin{subfigure}[b]{0.43\textwidth}
         \centering
         \includegraphics[width=\textwidth]{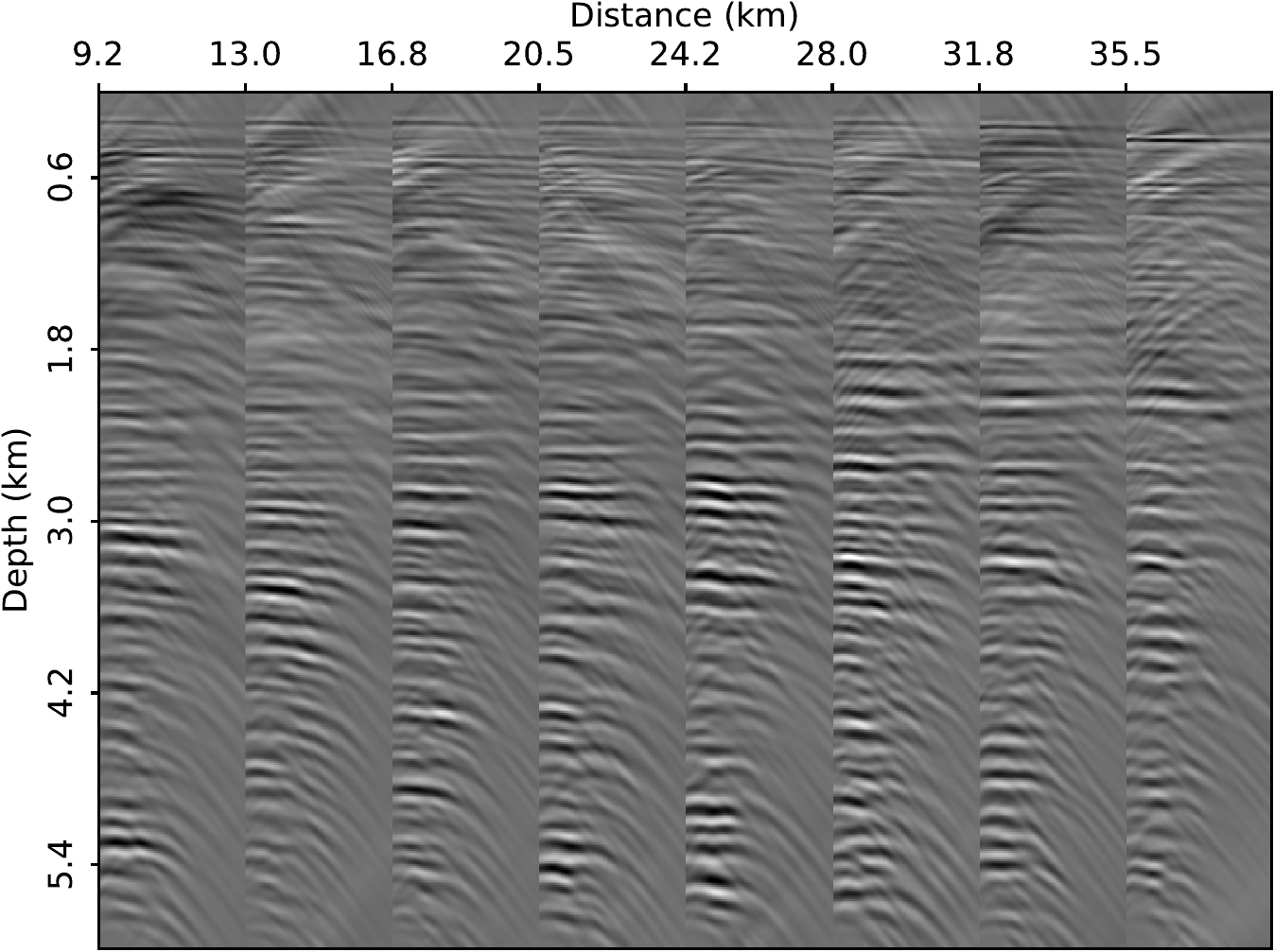}
         \caption{}
         \label{fig:three sin x}
     \end{subfigure}
      \caption{\label{fig:adcig012} Angle domain common image gathers (a) from the initial model and (b) from the final inverted model.}
\end{figure}

\begin{figure}[H]
     \centering
     \includegraphics[width=.43\textwidth]{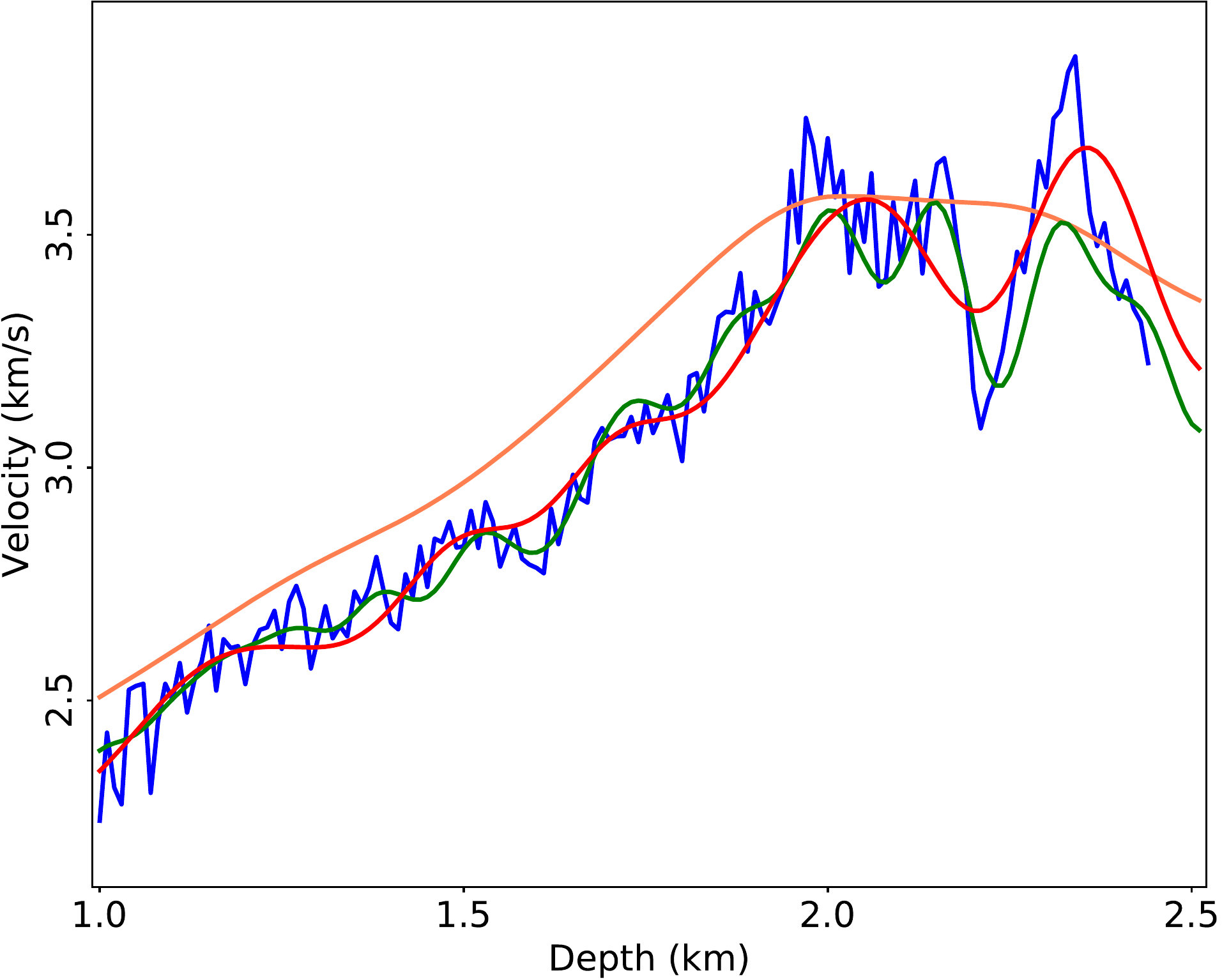}
     \caption{\label{fig:wlcmp} The P-wave velocity profiles comparison among the initial (coral), inverted from $\SI{3}{\hertz}-\SI{5}{\hertz}$ (red), inverted from $\SI{3}{\hertz}-\SI{10}{\hertz}$(green) and true model (blue). }
\end{figure}

Determining the optimal values for $\gamma$ and $\lambda$ is still an open question. As a rule of thumb, we choose $\gamma=10$ and $\lambda=99$ for the inversion of data in $\SI{3.0}{\hertz}$-$\SI{3.5}{\hertz}$ because this option does not visually present strong discontinuity in adjoint source as shown in Figure \ref{fig:adj99}c. Even though the combination of $\gamma=100$ and $\lambda=99$ used in Figure \ref{fig:adj99}d generates a smoother result, which could slow down the converge of wave-equation inversion. We use $\gamma=10$ and $\lambda=99$ for all four stages: $\SI{3.0}{\hertz}$-$\SI{3.5}{\hertz}$, $\SI{3.0}{\hertz}$-$\SI{5}{\hertz}$, $\SI{3.0}{\hertz}$-$\SI{7}{\hertz}$, and $\SI{3.0}{\hertz}$-$\SI{10}{\hertz}$. The observed data in the latter three stages are scaled by a constant such that the maximum amplitude is matched as in $\SI{3.0}{\hertz}$-$\SI{3.5}{\hertz}$. 
\begin{figure}[H]
     \centering
     \begin{subfigure}[b]{0.43\textwidth}
         \centering
         \includegraphics[width=\textwidth]{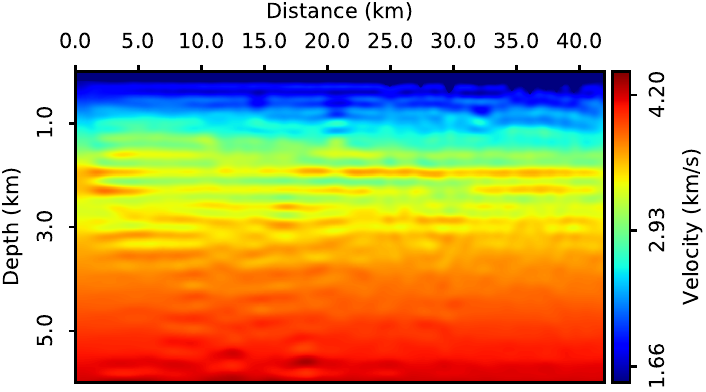}
         \caption{}
         \label{fig:y equals x}
     \end{subfigure}
     \begin{subfigure}[b]{0.43\textwidth}
         \centering
         \includegraphics[width=\textwidth]{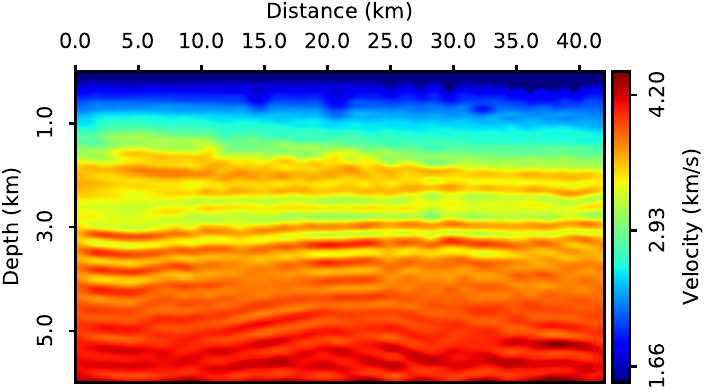}
         \caption{}
         \label{fig:three sin x}
     \end{subfigure}
      \caption{\label{fig:invert5} The inverted model (a) by mimetic non-differentiable ($\gamma=0.1$ and $\lambda=0$) and (b) penalized differentiable DTW ($\gamma=10$ and $\lambda=99$) misfit function from data in the frequency range $\SI{3}{\hertz}-\SI{5}{\hertz}$.}
\end{figure}
\begin{wrapfigure}{R}{9cm}
\centering
\includegraphics[width=.43\textwidth]{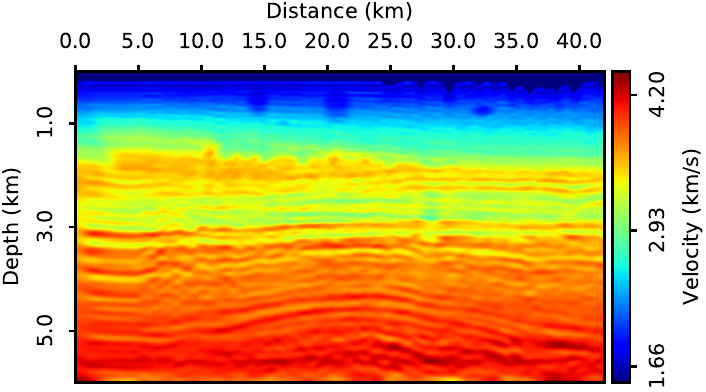}
    \caption{\label{fig:invert10} The inverted model by the penalized differentiable DTW misfit function from data in $\SI{3}{\hertz}-\SI{10}{\hertz}$.}
\end{wrapfigure}
The penalized differentiable DTW distance is equivalent to conventional nondifferentiable one in the case of $\gamma=0$ and $\lambda=0$ \cite[]{Cuturi2017SoftDTWAD}. From the above observation (Figure \ref{fig:adj0}, \ref{fig:adj9}, and \ref{fig:adj99}), we see a smaller $\gamma$ gives stronger abrupt changes in the adjoint source. Therefore we represent the nondifferentiable DTW distance with $\gamma=0.1$ and $\lambda=0$. Figure \ref{fig:invert5}a and \ref{fig:invert5}b shows the inverted model from data in $\SI{3.0}{\hertz}$-$\SI{5}{\hertz}$ by nondifferentiable DTW distance ($\gamma=0.1$ and $\lambda=0$) and penalized differentiable DTW distance ($\gamma=10$ and $\lambda=99$), respectively, where we see the result from non-differentiable DTW distance is not informative. Figure \ref{fig:invert10} is the inverted result by the penalized differentiable DTW distance with data in the frequency range $\SI{3.0}{\hertz}$-$\SI{10}{\hertz}$. 
Figure \ref{fig:rtm02} shows the migration result (a) from the initial and (b) from the inverted model, where we can see the continuity improves dramatically, especially for the deeper part. Figure \ref{fig:adcig012} shows the angle domain {[$0,\pi/4$]} common image gathers (a) from the initial and (b) from the inverted model, where we see the improvement from the velocity update, for example, the gathers at $x=\SI{24.2}{\kilo\meter}$ and $x=\SI{28.0}{\kilo\meter}$ are more flattened at larger reflection angles, especially in the depth from $\SI{1.8}{\kilo\meter}$ to $\SI{3.0}{\kilo\meter}$. Figure \ref{fig:wlcmp} shows the comparison of velocity profiles from the initial, inverted and true model, which illustrates that the model misfit is clearly reduced by wave-equation inversion based on the new misfit with $\gamma=10$ and $\lambda=99$.  
\section{Conclusion and Discussion}

The penalized differentiable DTW misfit function in its current form presents several weaknesses that are worth discussing. First of all, the calculation of the misfit value and its derivative is significantly expensive compared to the misfit based on $\mathcal{L}^2$-norm. We expect to develop an efficient algorithm based on the sparsity of the warping path. Besides, the misfit value from the new method can be negative. That brings some ambiguity to interpret the misfit history. Differentiable dynamic time warping divergence \citep{pmlr-v130-blondel21a} is one way to solve this problem; however, it leads to a considerable extra increase in the computational cost compared to the penalized differentiable misfit. If we employ the differentiable dynamic time warping divergence to solve the issue of negative misfit values, then a more efficient algorithm to calculate the DTW divergence and its derivative becomes urgent. Another weakness of the new misfit function is how to quantitatively determine the optimal values for smoothness and penalization parameters. 

We have introduced the penalized differentiable DTW misfit function for wave-equation inversion. There are two parameters in the proposed misfit function: the smoothness and penalization parameter. The former controls the smoothness of the warping path, and the latter determines how much the discrepancy information between the optimal warping plan and the expected one is added. A two-parameter inverse problem demonstrated that the penalized differentiable DTW misfit with proper smoothness and penalization parameters can present the expanded convex zone compared to the misfit functions based on non-differentiable DTW and the $\mathcal{L}^2$-norm. The inversion example with Chevron blind test data demonstrated that the new misfit function can resolve the presence of abnormally high amplitude in the adjoint source from the conventional non-differentiable DTW misfit. Thus the new misfit function improves the usability of DTW distance for wave-equation inversion. It also succeeded in retrieving a plausible velocity model from Chevron blind test data. 

\section*{Acknowledgments}
We acknowledge the Supercomputing Laboratory at King Abdullah University of Science \& Technology (KAUST) for providing resources that contributed to the research results reported within this manuscript. 
We would like to thank Chevron for making the SEG 2014 Chevron FWI synthetic available.

\newpage
\bibliographystyle{seg}  
\bibliography{example}

\end{document}